\newif\ifsingle

\ifsingle
\documentclass[12pt,draftclsnofoot, onecolumn]{IEEEtran}		
\else		
\documentclass[10pt,final, twocolumn]{IEEEtran}
\fi

\usepackage[T1]{fontenc}
\usepackage[latin9]{inputenc}
\usepackage{color}
\usepackage{amsmath}
\usepackage{amssymb}
\usepackage{graphicx}

\makeatletter

\providecommand{\tabularnewline}{\\}

\usepackage[noadjust]{cite}

\ifsingle
\newcommand{\figWidth}{0.65\columnwidth} 
\else
\newcommand{\figWidth}{0.9\columnwidth} 
\fi 

\definecolor{NewColor}{rgb}{0.2,0,0.5}

\makeatother

\begin{document}
\title{Deep Learning Based Successive Interference Cancellation for the Non-Orthogonal Downlink}
\author{Thien Van Luong, Nir Shlezinger,~\IEEEmembership{Member,~IEEE,} Chao Xu,~\IEEEmembership{Senior Member,~IEEE,} Tiep~M.~Hoang,~\IEEEmembership{Member,~IEEE,} Yonina C. Eldar,~\IEEEmembership{Fellow,~IEEE,} and~Lajos
Hanzo,~\IEEEmembership{Life Fellow,~IEEE}\thanks{This project has received funding from the European Union's Horizon
2020 research and innovation program under grant No. 646804-ERC-COG-BNYQ,
and from the Israel Science Foundation under grant No. 0100101. {L.~Hanzo} would like to acknowledge the financial support of the Engineering and Physical Sciences Research Council projects EP/P034284/1 and EP/P003990/1 (COALESCE) as well as of the European Research Council's Advanced Fellow Grant QuantCom (Grant No. 789028).} \thanks{T. V. Luong is with the Faculty of Computer Science, Phenikaa University, Hanoi 12116, Vietnam (e-mail: thien.luongvan@phenikaa-uni.edu.vn). }\thanks{
N. Shlezinger is with the School of ECE, Ben-Gurion University of
the Negev, Be'er-Sheva, Israel (e-mail: nirshl@bgu.ac.il).}  \thanks{C. Xu,  and L. Hanzo are with the School of Electronics and Computer Science,
University of Southampton, Southampton SO17 1BJ, U.K. (e-mail: cx1g08@ecs.soton.ac.uk,  lh@ecs.soton.ac.uk).} \thanks{T. M. Hoang was with the School of Electronics and Computer Science, University of Southampton, Southampton SO17 1BJ, U.K. He is now with the Department of Electrical Engineering, University of Colorado Denver, Denver, CO 80204, USA (e-mail: minhtiep.hoang@ucdenver.edu).} \thanks{
Y. C. Eldar is with the Faculty of Math and CS, Weizmann Institute
of Science, Rehovot, Israel (e-mail: yonina.eldar@weizmann.ac.il). } \vspace{-0.5cm}
 }

\maketitle
\vspace{-1cm}

\begin{abstract}
Non-orthogonal communications are expected to play a key role in
future wireless systems.  In downlink transmissions, the data symbols
are broadcast from a base station to different users, which are
superimposed with different power to facilitate high-integrity
detection using successive interference cancellation (SIC). However,
SIC requires accurate knowledge of both the channel model and channel state information (CSI), which may be difficult to acquire. We propose a deep learning-aided SIC
detector termed SICNet, which replaces the interference
cancellation blocks of SIC by deep neural networks (DNNs). Explicitly,
SICNet jointly trains its internal DNN-aided blocks for inferring the
soft information representing the interfering symbols in a data-driven
fashion, rather than using hard-decision decoders as in classical
SIC. As a result, SICNet reliably detects the superimposed symbols in
the downlink of non-orthogonal systems without requiring any prior
knowledge of the channel model, while being less sensitive to CSI
uncertainty than its model-based counterpart. SICNet is also robust to changes in the number of users and to their power allocation.
Furthermore, SICNet learns to produce accurate soft outputs, which
facilitates improved soft-input error correction decoding compared to
model-based SIC.  Finally, we propose an online training method for
SICNet under block fading, which exploits the channel decoding for
accurately recovering online data labels for retraining, hence,
allowing it to smoothly track the fading envelope without requiring
dedicated pilots. Our numerical results show that SICNet approaches
the performance of classical SIC under perfect CSI, while
outperforming it under realistic CSI uncertainty.
\end{abstract}


\vspace{-0.2cm}

\section{Introduction}

Wireless communications are facing escalating throughput, connectivity and scalability specifications. To meet these demanding
requirements, wireless systems may be expected to evolve
from conventional orthogonal to non-orthogonal solutions
\cite{Liu2017SurveyNOMA, Ding2017SurveyNOMA}. Non-orthogonal multiple access (NOMA) techniques allow users to simultaneously share the wireless channel resources for supporting heterogeneous end-devices, which inevitably imposes interference.

Sophisticated methods have been proposed for symbol detection in the
presence of interference \cite{andrews2005interference}.  In the
context of downlink (DL) non-orthogonal systems, where a base station (BS) transmits
a set of superimposed messages to different users over a shared
channel, the successive interference cancellation (SIC) algorithm has
been shown to be particularly suitable.  This is due to its
ability to approach the achievable rate region of such channels, when
combined with superposition coding at the BS \cite{Liu2017SurveyNOMA,Ding2017SurveyNOMA}, whilst its complexity only grows linearly with
the number of users.

The conventional SIC algorithm is model-based, i.e. it relies on
knowledge of the underlying statistical model. In particular,
implementing SIC detection requires each user to have accurate
knowledge of the channels between the BS and each of the users; 
its performance, however, is degraded in the presence of realistic imperfect
channel state information (CSI)
\cite{Yang2016imcsi}. Accurate estimation of CSI may be challenging, especially in rapidly fluctuating high-Doppler
frequency division duplexing scenarios, where the DL channels cannot
be estimated at the BS based on channel-reciprocity. Furthermore, the
conventional SIC algorithm assumes that the interference can be
cancelled by subtraction. However, this may not be the case in the presence of
non-linearities due to hardware impairments of low-resolution
analog-to-digital convertors  \cite{shlezinger2018asymptotic} and
non-linear amplifiers
\cite{iofedov2015mimo}. Finally, when the symbol detector has
to produce log-likelihood ratios (LLRs) for channel decoding, the SIC
algorithm typically suffers from model mismatch, since for simplicity it assumes Gaussian distributed residual interference, which has limited accuracy~\cite{kobayashi2001successive}.

An alternative approach to symbol detection, which does not rely on
any knowledge of the underlying channel model, is based on learning
the detection rule in a data-driven manner. There has been growing interest in the application of machine learning in digital
communication tasks, including symbol detection \cite{khani2020adaptive,Wang2020survey,farsad2020data}.
Deep neural networks (DNNs) are known to reliably infer knowledge in
complex environments \cite{Bengio09learning}, while relying solely on
data to learn their mapping. DNN-aided receivers can thus operate
accurately without requiring any knowledge of the underlying channel
model and its parameters.  Nonetheless, previous contributions on
DNN-based transceivers designed for the non-orthogonal DL, including
\cite{Luong2020MC-AE,Alberge2018Constell,Kang2020LearningSIC,Thien2020energy}, jointly learned the overall transmission path as an end-to-end autoencoder (AE). \textcolor{black}{For example, a  multicarrier autoencoder (MC-AE)  was proposed in \cite{Luong2020MC-AE}, which was shown to provide  enhanced frequency diversity gain for both coherent single-user and multi-user uplink (UL)/DL communications, outperforming its subcarrier index modulation based counterparts~\cite{Luong2018impact,Luong2018spread}. A similar MC-AE concept was introduced for energy detection-based non-coherent systems  in~\cite{Thien2020energy}. These previous contributions assume an equal power allocation scheme applied to all users, therefore, limiting their suitability for exploiting the established benefits of SIC detection combined with superposition coding. In \cite{Kang2020LearningSIC}, a precoder and an SIC-based decoder were jointly optimized as an AE in the multi-input multi-output NOMA (MIMO-NOMA) DL, assuming the availability of perfect CSI at the transmitter side. Finally, the constellation of a two-user NOMA DL was designed by training an AE in~\cite{Alberge2018Constell}. However, all these AE-based schemes require knowledge of the channel model for jointly training both the transmitter and receiver, and - similarly to the classic model-based techniques - require accurate CSI.}

Conventional DNNs require massive amounts of data for training, and
lack the clear physical interpretation of model-based approaches.  It
was recently proposed to integrate DNNs into model-based symbol
detection
algorithms \cite{Shlezinger2020ViterbiNet,shlezinger2020data,Shlezinger2020DeepSIC},
resulting in hybrid model-based/data-driven receivers, which learn to
carry out established detection algorithms from relatively small data
sets without requiring any knowledge of the channel model. In
particular, the authors
of \cite{Shlezinger2020ViterbiNet,shlezinger2020data} introduced
data-driven implementations of both the Viterbi algorithm and of the
BCJR scheme, which are applicable for finite-memory
channels. \textcolor{black}{Furthermore, the authors of \cite{Shlezinger2020DeepSIC} presented a receiver that learns to carry out soft interference cancellation. \textcolor{black}{This receiver operation is designed for the UL of non-orthogonal systems, where the task is to detect all transmitted symbols. By contrast, in the non-orthogonal downlink of this treatise, the receiver only has to recover its corresponding symbol that is corrupted by interference}.}  These previously proposed DNN-aided symbol detectors motivate the design of a hybrid model-based/data-driven implementation of the SIC algorithm for the DL of non-orthogonal
systems, which is our focus here.

\begin{table*}[ht]
\centering \caption{Comparing our contributions to the literature of learning for the DL of non-orthogonal systems}
\label{tab:contribution} \linespread{1.0} %
\begin{tabular}{|l|c|c|c|c|c|}
\hline 
Contribution  & \cite{Luong2020MC-AE}  & \cite{Alberge2018Constell}  & \cite{Kang2020LearningSIC}  & \cite{Thien2020energy}  & This work\tabularnewline
\hline 
\hline 
Hybrid model-based and data-driven designs  &  &  & \checkmark  &  & \checkmark\tabularnewline
\hline 
End-to-end autoencoder-based designs  & \checkmark  & \checkmark  & \checkmark  & \checkmark  & \tabularnewline
\hline 
No dedicated pilot transmissions  &  &  &  & \checkmark  & \checkmark\tabularnewline
\hline 
Independent of channel modelling  &  &  &  &  & \checkmark\tabularnewline
\hline 
Small dataset (few thousands of samples)  &  &  &  &  & \checkmark\tabularnewline
\hline 
Robustness to superposition coding perturbations &  &  &  &  & \checkmark\tabularnewline
\hline 
FEC-aided soft decoding  &  &  &  &  & \checkmark\tabularnewline
\hline 
Online self-supervised learning  &  &  &  &  & \checkmark\tabularnewline
\hline 
Imperfect CSI  & \checkmark  &  &  &  & \checkmark\tabularnewline
\hline 
Non-linear channels  &  &  &  &  & \checkmark\tabularnewline
\hline 
\end{tabular}
\end{table*}

	\textcolor{black}{In this contribution, we present SICNet, which is a DNN-aided receiver architecture that learns to implement the SIC algorithm from labeled data. SICNet
is derived by representing the SIC algorithm as an interconnection
of basic building blocks, each trained to cancel the interference
imposed by a given user. Despite the similar acronym, SICNet is fundamentally different from DeepSIC \cite{Shlezinger2020DeepSIC}. Although both receivers belong to the class of hybrid model-aided networks \cite{shlezinger2020model}, they differ both in their objective and in their operation. Specifically, SICNet is designed for the non-orthogonal DL, where the task is to recover a single desired symbol in the presence of both interference as well as noise, and does so by learning to implement the SIC algorithm, which is known to be eminently suitable for such scenarios. By contrast, DeepSIC focuses on the joint recovery of multiple interfering symbols, representing an UL setup, while relying on the classic parallel soft interference cancellation method \cite{choi2000iterative}. Thus, the scheme in \cite{Shlezinger2020DeepSIC} relies on a larger number of detection and interference cancellation steps compared to SIC, since the goal of SIC is to detect a single symbol \cite{andrews2005interference}. As a result, the overall architecture of SICNet is different from that of DeepSIC, and it harnesses a much lower number of neural building blocks, making it more suitable for mobile DL receivers.} 

\textcolor{black}{Once trained, SICNet implements SIC detection, without requiring
any knowledge of the underlying channel model, e.g.,
without restricting the operation to linear channels. We demonstrate that SICNet trained on data from a channel
with a given signal-to-noise ratio (SNR) approaches the performance
of the model-based SIC algorithm used for symbol detection, which relies on accurate SNR-dependent CSI. Furthermore, SICNet substantially
outperforms its model-based counterpart in the presence of CSI uncertainty, under both linear and non-linear channels,
indicating its potential to facilitate accurate symbol
detection in non-orthogonal DL systems. Additionally, SICNet can readily adapt to time-variant DL scenarios, such as adding a new user and changing the order of the power assignment among the users, at the cost of low-complexity retraining and without requiring to rebuild its DNN structure.}

\textcolor{black}{We also show that, when SICNet is used for producing soft symbols provided for a forward error correction (FEC) decoder, it
yields improved decoding accuracy compared to using the model-based
SIC with full CSI for the same purpose. This is a benefit of the fact that
SICNet, which operates in a model-agnostic manner, learns to compute
more accurate bit-wise LLRs compared
to SIC, which assumes a Gaussian distributed interference recovery error. 
Finally, we design an FEC coding-aided online training method for
SICNet in order to make its DNNs adapt to the variations of block
fading channels without requiring new training data. In particular,
we exploit the presence of FEC codes as indication for
the correctness in detecting a block of symbols, as done in \cite{Shlezinger2020ViterbiNet,teng2020syndrome, raviv2021meta},
in order to accurately form a relatively small number of labels, which are sufficient for retraining SICNet
with a few epochs. Table~\ref{tab:contribution} summarizes the main contributions of this work and explicitly compares them to the literature of learning-aided DL detection in non-orthogonal systems.}

The rest of this paper is organized as follows. Section~\ref{sec:System-Model}
details the system model and briefly reviews the SIC algorithm. Section~\ref{sec:SICNet}
presents the proposed SICNet. Section~\ref{sec:online-training}
discusses how SICNet can be combined with FEC decoding and FEC-aided
online training. Our numerical evaluations are provided in Section~\ref{sec:Numerical-Evaluations}.
Finally, Section~\ref{sec:Conclusions} concludes the paper.

Throughout the paper, $\mathbb{R}$ denotes the set of real numbers,
and $\mathbb{R}^{n}$ stands for the $n$ Cartesian product of $\mathbb{R}$.
We use $\mathbb{E}\left[\cdot\right]$, $p(\cdot)$, and $\Pr(\cdot)$
for the stochastic expectation, probability density function (PDF),
and probability mass function, respectively, while $\mathcal{N}\left(0,\sigma^{2}\right)$
is the Gaussian distribution with zero mean and variance $\sigma^{2}.$


\vspace{-0.2cm}

\section{System Model\label{sec:System-Model}}

\vspace{-0.1cm}
We begin by describing the system model for which we derive SICNet. With
that aim, we first present the channel model in Subsection \ref{subsec:NOMA-Channel-Model}
and then formulate the symbol detection problem in Subsection
\ref{subsec:Problem-Formulation}. We then review the model-based
SIC algorithm in Subsection \ref{subsec:Successive-Inteference-Cancellat}.

\vspace{-0.2cm}

\subsection{Non-orthogonal DL Channel Model\label{subsec:NOMA-Channel-Model}}

\begin{figure}[tb]
\begin{centering}
\includegraphics[width=\figWidth]{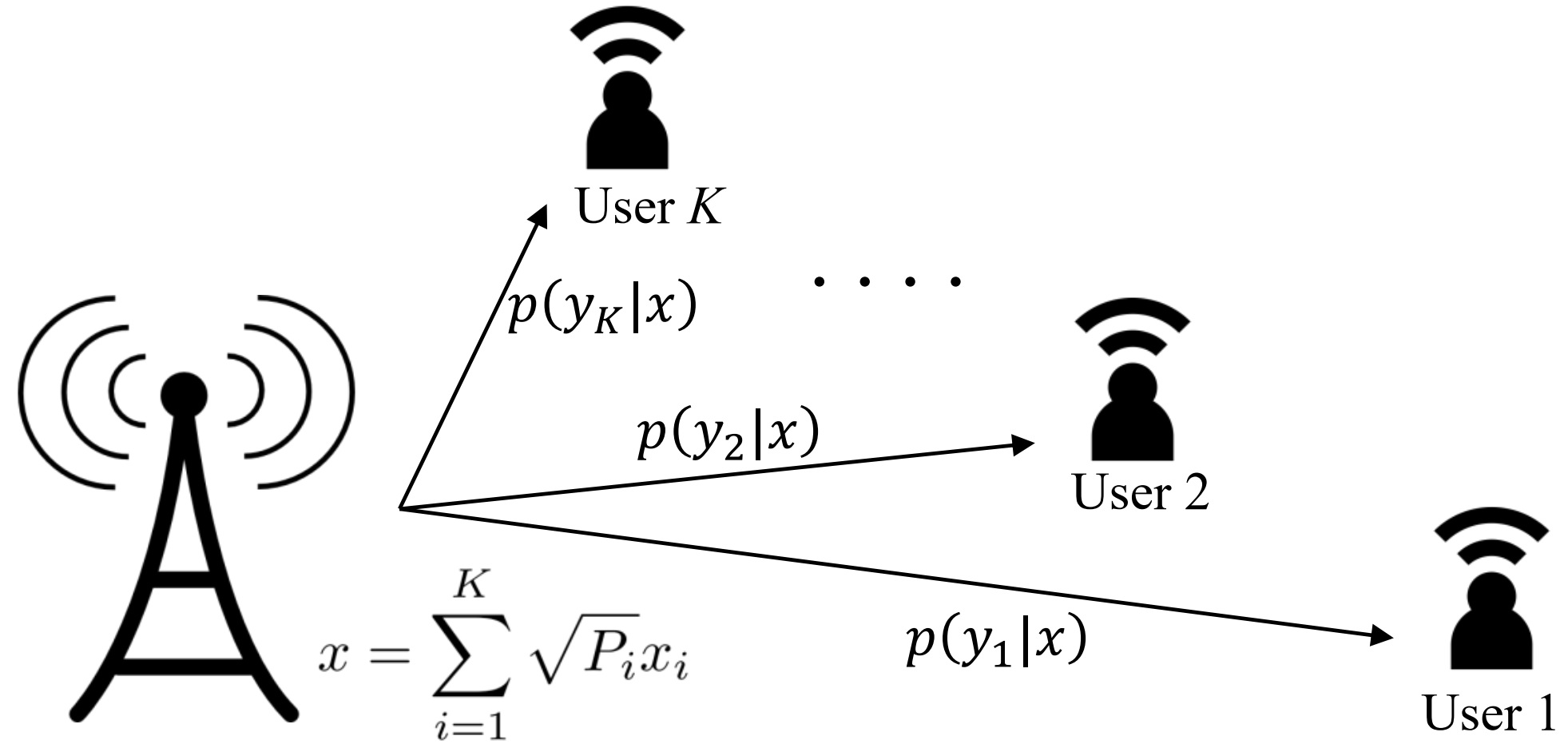} 
\par\end{centering}
\caption{Downlink non-orthogonal system with $K$ users.\label{fig:noma}}
\end{figure}

\vspace{-0.1cm}
 Consider a non-orthogonal DL, where a BS transmits
data simultaneously to $K$ users within the same time- and frequency-
resources, as illustrated in Fig. \ref{fig:noma}. For simplicity,
we focus on scenarios where both the BS and the users are equipped
with a single antenna. The BS transmits a set of symbols $\{x_{k}\}_{k=1}^{K},$
each intended for a different user, via superposition coding, as in the power-domain non-orthogonal DL \cite{Ding2017SurveyNOMA}. In particular,
the symbol $x_{k}$ intended to user $k$ is amplified with transmitted
power $P_{k}$ for $k=1,...,K$. These signals are superimposed at
the BS, resulting in the channel input $x$ which is given by: 
\begin{equation}
x=\sum_{k=1}^{K}\sqrt{P_{k}}x_{k}.\label{eq:x_BS}
\end{equation}

\textcolor{black}{We assume that the symbols are mutually independent, and that each
symbol $x_{k}\in\mathbb{R}$ is drawn from an $M$-point constellation
$\mathcal{S},$ having unit mean power, i.e., $\left|\mathcal{S}\right|=M$, and $\mathbb{E}\big[|x_{k}|^{2}\big]=1$. While the digital constellation is assumed to have unit power regardless of its order $M$, the superposition coding utilized in the downlink  scales the power of each transmitted symbol via the coefficients $\{P_k\}$, to facilitate decoding, as detailed in  \cite{Liu2017SurveyNOMA,Ding2017SurveyNOMA}.
For the sake of simplicity, we assume that the symbols have the same modulation order $M$, although it is straightforward to extend our work to a generalized scenario, where different modulation orders are used for different users. Our work can also be easily 
adapted to complex-valued signals, by representing them using real vectors of extended dimension.} 

While we do not impose a specific model on the DL channel, we assume
that it is memoryless and that the channel outputs at the $K$ users,
denoted $\{y_{k}\}_{k=1}^{K},$ are mutually independent conditioned
on $x$, i.e. the joint conditional PDF of the channel outputs satisfies
\begin{equation}
p(y_{1,}\ldots,y_{K}|x)=\prod_{k=1}^{K}p(y_{k}|x).\label{eq:cond_pdf}
\end{equation}
A commonly used DL model which obeys \eqref{eq:cond_pdf} is
the linear Gaussian broadcast channel. Here, the channel output observed
by user $k$ is given by 
\begin{equation}
y_{k}=h_{k}x+w_{k}=h_{k}\left(\sum_{i=1}^{K}\sqrt{P_{i}}x_{i}\right)+w_{k},\label{eq:y_k}
\end{equation}
where $h_{k}\in\mathbb{R}$ is the channel coefficient between the
BS and user $k$, and $w_{k}\in\mathbb{R}$ is  additive white
Gaussian noise (AWGN).

\vspace{-0.2cm}

\subsection{Problem Formulation\label{subsec:Problem-Formulation}}

\vspace{-0.1cm}
 Our goal is to design a symbol detection mechanism for each user
of index $k=1,...,K$, namely, a mapping $\hat{x}_{k}:\mathbb{R\mapsto\mathcal{S}},$
so that $\hat{x}_{k}$ is an estimate of $x_{k}$ from the observed
channel output $y_{k}$. As detailed in the previous subsection, we
do not assume any prior knowledge of the channel model at the receiver,
except that its input--output relationship takes the generic form
in \eqref{eq:cond_pdf}. \textcolor{black}{Furthermore, we do not require the users to know their power allocation coefficients $\{P_{k}\}_{k=1}^{K}$,
but we assume that they know their order, which is written henceforth
as $P_{1}\ge P_{2}...\ge P_{K}$ without loss of generality. Note that the conventional SIC requires that each user knows both the power of all users and their power order, in addition to accurate channel knowledge.} Each
user of index $k$ has access to a labeled data set of $T$ samples,
denoted by $\{y_{k}^{(t)},x_{1}^{(t)},\ldots,x_{K}^{(t)}\}{}_{t=1}^{T}.$
In practice, such data typically corresponds to preamble and pilot
transmissions. We assume that the number of pilots is limited to be
on the order of a few several thousands of samples, which is the length
of a typical LTE preamble \cite[Ch. 17]{dahlman20103g}. 

The lack of channel model knowledge combined with the presence of
labeled data motivates a data-driven design based on DNNs. However,
the fact that the dataset is limited, indicates that it is preferable
to incorporate some domain knowledge in our design, rather than directly applying
a black-box DNN. In particular, the relevant domain knowledge
here is that in  the downlink, the symbols are mutually independent,
take values in $\mathcal{S}$ and are superimposed with power allocations
satisfying $P_{1}\ge P_{2}...\ge P_{K}$. For such scenarios, it is
preferable for the $k$-th user to successively detect the interfering
symbols $x_{1},\ldots,x_{k-1}$ before recovering its desired $x_{k}$,
rather than detecting it directly. This recovery mechanism is the
SIC algorithm, detailed in the following.

\vspace{-0.2cm}

\subsection{Successive Interference Cancellation\label{subsec:Successive-Inteference-Cancellat}}

\vspace{-0.1cm}
 The SIC algorithm is commonly adopted in the NOMA literature, due
to its simplicity and its ability to approach the achievable rate
region of linear Gaussian non-orthogonal broadcast channels \eqref{eq:y_k}, when combined
with superposition coding \cite{Liu2017SurveyNOMA}. \textcolor{black}{A superposition code determines the power assigned to the symbol intended for each user. A common approach to select these codes is to allocate more power to users having poorer channel gains \cite{Liu2017SurveyNOMA,Ding2017SurveyNOMA,Yuan2018codedSIC}. Such a formulation, which is intended  to facilitate detection at each user and boost fairness, requires some assessment of the quality-based ordering of the individual channels at each receiver. Alternatively, one can determine the superposition code based on the application layer requirements and priorities. Regardless of how the superposition code is determined, it controls the power levels $\{P_k\}_{k=1}^{K}$, and we henceforth assume that  $P_{1}\ge P_{2}...\ge P_{K}$.} 

To formulate the model-based SIC algorithm,  consider a linear Gaussian channel
\eqref{eq:y_k}.
Based on this, the SIC detector of user $k$ operates in the following
iterative fashion. First, user $k$ detects the signal of the user having the highest power, i.e., user 1, while treating the interference
as noise, using the maximum likelihood (ML) criterion, which here
is given by $\hat{x}_{1}=\arg\min_{x_{1}\in\mathcal{S}}\left|y_{k}-\sqrt{P_{1}}h_{k}x_{1}\right|.$ Then, the contribution of user 1 to $y_{k}$ is eliminated for decoding
the signal of user 2. Explicitly, the symbol of user 2 is recovered using
the ML estimate in which the interfering signal of user 1 is estimated
by $\hat{x}_{1}$, yielding, 
\begin{equation}
\hat{x}_{2}=\arg\min_{x_{2}\in\mathcal{S}}\left|\left(y_{k}-\sqrt{P_{1}}h_{k}\hat{x}_{1}\right)-\sqrt{P_{2}}h_{k}x_{2}\right|.\label{eq:sic_user_2}
\end{equation}
This SIC process continues in this manner recursively, until the symbol
of user $k$ is detected. This can be achieved by hard decision, i.e.,
providing an estimate of the transmit ${x}_{k}$ via 
\begin{equation}
\hat{x}_{k}=\arg\min_{x_{k}\in\mathcal{S}}\left|\left(y_{k}-\sum_{i=1}^{k-1}\sqrt{P_{i}}h_{k}\hat{x}_{i}\right)-\sqrt{P_{k}}h_{k}x_{k}\right|.\label{eq:sic_user_k}
\end{equation}

	\textcolor{black}{The usage of different power assigned to different users allows user $k$ to detect the symbols of its preceding users, namely $x_{1},...,x_{k-1}$, with high accuracy. This makes the SIC procedure particularly suitable for its symbol detection in the non-orthogonal downlink at a low complexity and high reliability. For comparison, if user $k$ directly detects its own symbol while treating the signals of other users as interference, it is likely to achieve degraded detection performance due to the presence of severe interference from other users, which SIC cancels by its iterative procedure.}  

Alternatively, SIC can be used to provide soft outputs represented by
the LLR for each bit embedded in the symbol ${x}_{k}$. These outputs
are particularly useful when combined with soft-input FEC decoders.
In particular, letting $\beta_{n}$ be the $n$-th bit of symbol $x_{k}$,
we partition $\mathcal{S}$ into two subsets $\mathcal{S}_{n}^{(0)}$
and $\mathcal{S}_{n}^{(1)}$ which satisfy $\beta_{n}=0$ and $\beta_{n}=1$,
respectively, i.e., $\mathcal{S}_{n}^{(0)}\cup\mathcal{S}_{n}^{(1)}=\mathcal{S}$. Here, we assume that user $k$ does not know the coding schemes of other users, i.e. its FEC decoder does not decode the transmitted bits of other users for SIC operation, but directly decodes its own bits only. As such,  
upon denoting $z=y_{k}-\sum_{i=1}^{k-1}\sqrt{P_{i}}h_{k}\hat{x}_{i}$,
 \textcolor{black}{when the constellation symbols are equiprobable}, the LLR of $\beta_{n}$ can be expressed from \eqref{eq:sic_user_k} as 
\begin{equation}
L_{n}=\text{log}\frac{\text{Pr}\left(\beta_{n}=0|z\right)}{\text{Pr}\left(\beta_{n}=1|z\right)}=\log\frac{\sum_{x_{k}\in\mathcal{S}_{n}^{(0)}}p\left(z|x_{k}\right)}{\sum_{x_{k}\in\mathcal{S}_{n}^{(1)}}p\left(z|x_{k}\right)},\label{eq:LLR_bits_SIC}
\end{equation}
where $p\left(z|x_{k}\right)$ is the PDF of $z$ conditioned on $x_{k}$.
Note that it is difficult to exactly determine $p\left(z|x_{k}\right)$.
Therefore, in order to estimate the LLR $L_{n}$, the interference-detection-error-plus-noise
term $z-\sqrt{P_{k}}h_{k}x_{k}$ is often approximated by Gaussian
noise $w_{k}$ in \eqref{eq:y_k} with zero mean and variance $\sigma^{2}$,
resulting in \cite{Yuan2018codedSIC} 
\begin{equation}
p\left(z|x_{k}\right)\approx\frac{1}{\sqrt{2\pi}\sigma}\exp\left(-\left|z-\sqrt{P_{k}}h_{k}x_{k}\right|^{2}/2\sigma^{2}\right).\label{eqn:PDFApprox}
\end{equation}

The combination of superimposed coding and SIC detection allows the BS
to simultaneously serve multiple users with the same resources, while
achieving significantly improved bandwidth efficiency over orthogonal
architectures. However, in order to implement SIC in the
non-orthogonal DL, the receiver must have exact CSI for each user,
i.e. evaluating \eqref{eq:sic_user_k} requires accurate knowledge of
$h_{k}$. \textcolor{black}{In particular, the detection performance of SIC strongly
depends on the accuracy of recovering the interfering symbols in the
preceding iterations. There is a significant performance loss, when the
CSI of the users is imperfect, as shown in \cite{Yang2016imcsi}. In some important wireless scenarios, including rapidly fluctuating high-Doppler frequency division duplexing scenarios and the family of systems aided by reconfigurable intelligent surfaces \cite{alexandropoulos2021reconfigurable}, obtaining accurate CSI may be challenging. Another limitation is that the channel has to obeys the linear form
in \eqref{eq:y_k}, for which the detector can cancel the interference
by demodulation, remodulation and subtraction, making it suitable only for linear
channels. Such models may not hold when using low-resolution receivers \cite{shlezinger2018asymptotic} and non-linear amplifiers \cite{iofedov2015mimo}.}  Each user is also required to know the power
allocation coefficients assigned to each of the users in the network
for reliable symbol detection. Moreover, when soft outputs are
required, the SIC may be unable to provide an accurate estimate of the
LLRs to be used by a soft FEC decoder due to the approximation of the
conditional PDF in \eqref{eqn:PDFApprox} as being Gaussian. 

	\textcolor{black}{Such fundamental limitations of the SIC, combined with the feasibility of integrating DNNs into model-based receiver algorithms for learning-aided computation of specific model-based steps \cite{Shlezinger2020ViterbiNet,shlezinger2020data} including interference cancellation \cite{Shlezinger2020DeepSIC} motivates the use of DNNs to replace the interference cancellation blocks of SIC. This allows continued operation, when the knowledge of accurate CSI, the channel model, the power coefficients and even $M$-ary modulation type are no longer required at the user side, as presented in the next sections.}

\vspace{-0.2cm}

\section{SICNet\label{sec:SICNet}}

\vspace{-0.1cm}
 To address the aforementioned issues of the conventional SIC
receiver, we propose a DNN-based SIC detector called
SICNet. Explicitly, SICNet uses deep learning to recover a soft estimate of the
interference of each user rather than applying the hard-decision ML
detector used in the conventional scheme. In the following, we present
the architecture of our SICNet in Subsection \ref{subsec:SICNet-Architecture},
followed by its training procedure and a discussion in Subsections
\ref{subsec:Training-SICNet}-\ref{subsec:Discussion}.

\vspace{-0.2cm}

\subsection{SICNet Architecture\label{subsec:SICNet-Architecture}}

\vspace{-0.1cm}

\begin{figure}[tb]
\begin{centering}
\includegraphics[width=\figWidth]{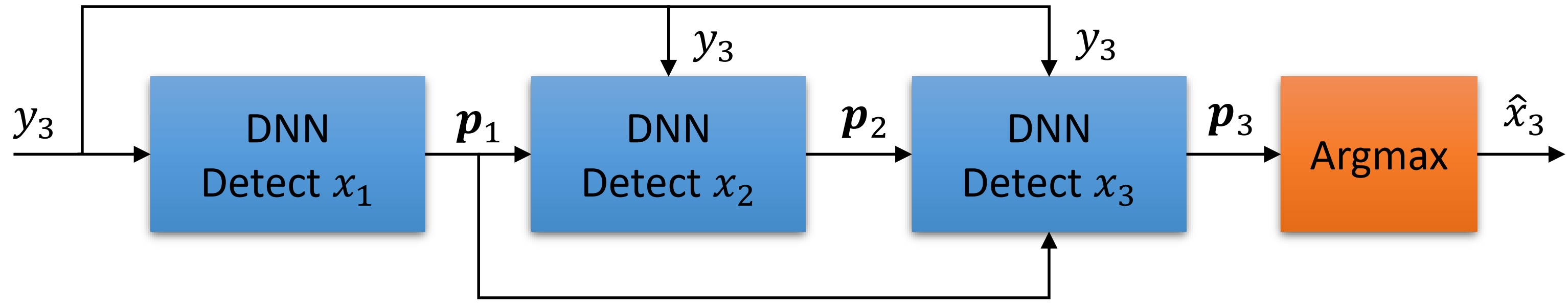} 
\par\end{centering}
\caption{Proposed SICNet detector of user 3 in a non-orthogonal downlink channel shared by  $K=3$ users.\label{fig:SICNet}}
\end{figure}

The architecture of SICNet is illustrated in Fig. \ref{fig:SICNet}.
For the sake of simplicity, we consider a  non-orthogonal DL supporting $K=3$
users, and focus our description of the architecture on user $k=3$.
A SICNet architecture designed for $K$ users
can be devised based on Fig. \ref{fig:SICNet} described as follows.

Our design of SICNet builds upon the insight that the SIC method is
comprised of multiple basic building blocks, each corresponding to the recovery
of the symbol of a different user. Inspired by \cite{Shlezinger2020DeepSIC}, we implement SIC in a data-driven fashion by preserving its overall flow as an interconnection of building blocks, while replacing each block by a dedicated DNN. In particular, each building block implements symbol recovery, and can thus be treated as a classification task, which is capable of learning from data in a
model-agnostic manner using deep classifiers. As a result, SICNet of user $k$ consists of $k$ different DNN blocks,
where DNN block $i$ is used to detect the soft information $\mathbf{p}_{i}$ of user
$i$ for $i=1,...,k$. More particularly,
$\mathbf{p}_{i}\in\mathbb{R}^{M}$ represents an estimate of the conditional
distribution of the corresponding symbol, given the past estimates, formulated as:
\begin{equation}
\mathbf{p}_{i}=\left[\begin{array}{c}
\hat{p}\left(x_{i}=\alpha_{1}|y_{k},\mathbf{p}_{1},\ldots,\boldsymbol{\mathbf{p}}_{i-1}\right)\\
\vdots\\
\hat{p}\left(x_{i}=\alpha_{M}|y_{k},\mathbf{p}_{1},\ldots,\boldsymbol{\mathbf{p}}_{i-1}\right)
\end{array}\right],\label{eq:p_i}
\end{equation}
where $\alpha_{j}$ is the $j$-th constellation symbol of $\mathcal{S}$
and $\hat{p}\left(x_{i}=\alpha_{j}|y_{k},\mathbf{p}_{1},\ldots,\boldsymbol{\mathbf{p}}_{i-1}\right)$
is a parametric estimate of the probability of $x_{i}$ decoded as
$\alpha_{j}$ conditioned on $y_{k}$ and the previous soft estimates
$\mathbf{p}_{1},\ldots,\boldsymbol{\mathbf{p}}_{i-1}$, for $j=1,...,M$.

\begin{figure}
\centering \includegraphics[width=\figWidth]{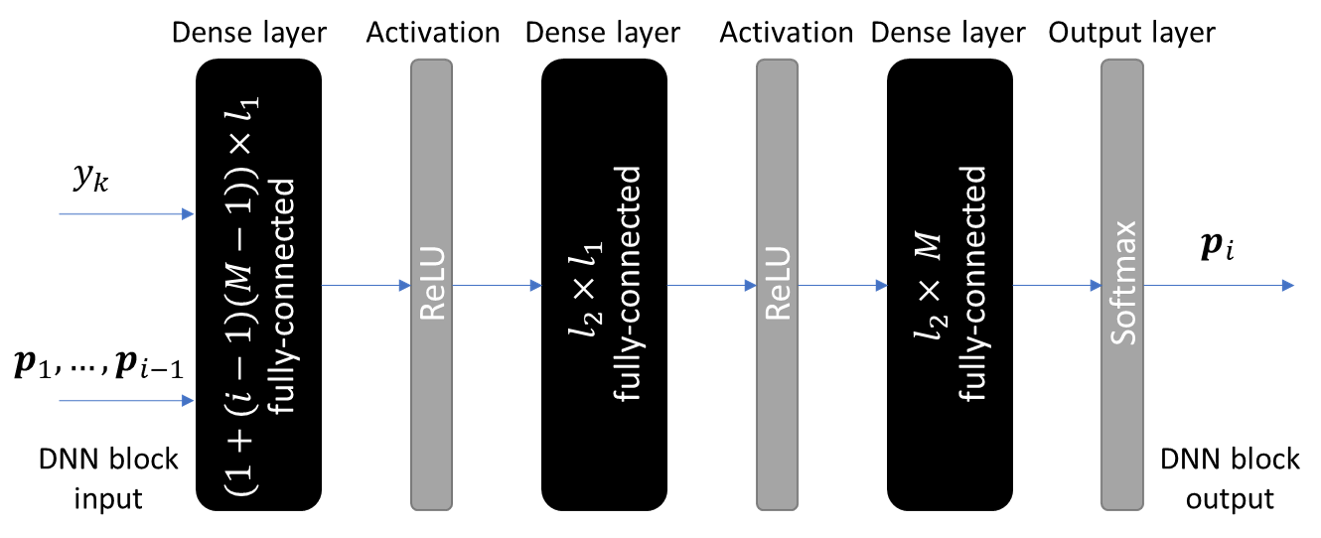}
\caption{An illustration of the $i$-th DNN of SICNet.}
\label{fig:DNN_arch1} 
\end{figure}

In SICNet, each conditional distribution estimate $\mathbf{p}_{i}$
is the output vector of the DNN block $i$, which satisfies $\sum_{j=1}^{M}\hat{p}\left(x_{i}=\alpha_{j}|y_{k},\mathbf{p}_{1},\ldots,\boldsymbol{\mathbf{p}}_{i-1}\right)=1.$
This can be naturally implemented by using a softmax activation \cite{Thien2020energy}
at the output layer of each DNN block. The input data of DNN block
$i$ includes both $y_{k}$ and the outputs from $i-1$ former blocks,
namely $\mathbf{p}_{1},...,\mathbf{p}_{i-1}$. More specifically,
those elements are concatenated to form an input vector of the size
$\left[1+\left(i-1\right)M\right]$ for DNN block $i$, which can
be reduced to length $\left[1+\left(i-1\right)(M-1)\right]$, since
the sum of the entries of each $\mathbf{p}_{i}$ always equals one.
Thus its last entry is determined by its first $M-1$ entries. An
illustration of an implementation of the $i$-th building block DNN
using a fully-connected network having two hidden layers, as used in
our numerical study in Section~\ref{sec:Numerical-Evaluations},
is depicted in Fig.~\ref{fig:DNN_arch1}.  \textcolor{black}{As seen in Fig.~\ref{fig:SICNet}, the input to SICNet, which is the input of the first DNN block ($i=1$), is  $y_{k}$. As such, the input of SICNet for user $k$ is only its received signal $y_{k}$, i.e., no CSI information and no prior knowledge of the power allocation $\{P_k\}$ is required at each user.}

 Finally, following \eqref{eq:sic_user_k}, a hard estimate
of the symbol of interest $x_{k}$ is obtained by taking to the largest
element of $\mathbf{p}_{k},$ which is the output vector of DNN block
$k$, i.e., 
\begin{equation}
\hat{x}_{k}=\arg\max_{\alpha\in\mathcal{S}}\hat{p}\left(x_{k}=\alpha|y_{k},\mathbf{p}_{1},\ldots,\boldsymbol{\mathbf{p}}_{k-1}\right).\label{eq:x_hat}
\end{equation}
Furthermore, SICNet can also be used to provide bit-wise LLR estimates, as we will discuss in Section~\ref{sec:online-training}.

While the design of SICNet is inspired by DeepSIC, proposed in \cite{Shlezinger2020DeepSIC} for multi-user detection in the non-orthogonal uplink, the resultant model-aided networks are notably different. In particular, the number of neural building blocks in SICNet is determined by the order of the specific user in the superposition code, as illustrated in Fig.~\ref{fig:SICNet}. For instance, the architecture of the receiver of user $2$ is comprised of $2$ DNNs blocks, while user $3$ utilizes $3$ such blocks. Nonetheless, SICNet can also cope with perturbations of the order of the users in the superposition code without having to change its architecture, as we numerically demonstrate in Subsection~\ref{subsec:SERRes}. DeepSIC requires a much larger number of building blocks, which is set to the number of users in the UL, multiplied by a fixed number of iterations, typically $5$. Furthermore, the successive operation of SICNet implies that each constituent DNN has a different number of inputs,  as illustrated in Fig.~\ref{fig:DNN_arch1}, while in DeepSIC the architecture of all constituent DNNs is identical, since each building block takes the soft-detection representation of all interfering symbols as its inputs. Consequently, SICNet uses a small number of neural building blocks compared to DeepSIC, and each DNN block differs from that used by DeepSIC.

In contrast to the conventional SIC, SICNet uses a soft estimate of the interfering
symbols, which is not hard-canceled by subtraction,
hence it is not restricted to channels of the form \eqref{eq:y_k}.
Furthermore, the model-agnostic nature of DNN classifiers and their
ability to operate reliably in complex and analytically intractable
settings imply that SICNet does not require the knowledge of the channel model in
its detection process. Hence, our scheme can work for arbitrary channel
models in a data-driven manner, which is not the case for its  classical
counterpart. Finally, SICNet does not require its users to know the power coefficients of other users and their modulation schemes, while classical SIC relies on this information in its detection process, as shown in \eqref{eq:sic_user_2}-\eqref{eq:sic_user_k}. In fact, SICNet only has to know the rank-order of user powers and the modulation alphabet size $M$, which decide the number of DNN blocks and the output dimension of each DNN block, respectively.

\vspace{-0.2cm}

\subsection{Training SICNet\label{subsec:Training-SICNet}}

\vspace{-0.1cm}
 Next, we describe the training procedure of SICNet, focusing on an
arbitrary user of index $k$. First, we represent the training data
as $\big\{ y_{k}^{(t)},\mathbf{q}_{1}^{(t)},...,\mathbf{q}_{k}^{(t)}\big\}$,
where $y_{k}$ is the received signal of user $k$, and $\mathbf{q}_{i}\in\mathbb{R}^{M}$
is the one-hot encoding of $x_{i}$, for $i=1,...,k$, representing
the true label of $\mathbf{p}_{i}$, i.e., the output of DNN block
$i$. As $\mathbf{q}_{i}$ is a one-hot vector, its elements are all
zeros, except for a unique element being one. The index of this unit
element is the index of the constellation symbol of user $i$, which
is $m$ for $x_{i}=\alpha_{m}\in\mathcal{S}$. \textcolor{black}{Using the softmax activation as the  output  layer  of  each  DNN  block  in  SICNet produces a soft probabilistic estimate of the corresponding symbol. Consequently, the loss measure is based on the cross entropy function, which is a well-established loss function for training deep classifiers amongst others, because it facilitates gradient based training \cite{goodfellow2016deep}.} 
The resulting loss is computed over each batch of $T$ data samples as follows:
\begin{equation}
\mathcal{L}\left(\boldsymbol{\theta}\right)=-\frac{1}{T}\sum_{t=1}^{T}\sum_{i=1}^{k}\varphi_{i}\sum_{j=1}^{M}q_{i,j}^{(t)}\log p_{i,j}^{(t)},\label{eq:loss}
\end{equation}
where $\boldsymbol{\theta}$ denotes the trainable parameters of SICNet
including the weights and biases of all DNN blocks, $q_{i,j}^{(t)}$ and
$p_{i,j}^{(t)}$ are the $j$-th elements of $\mathbf{q}_{i}^{(t)}$
and $\mathbf{p}_{i}^{(t)}$, respectively, where $\mathbf{p}_{i}^{(t)}$
is the output of DNN block $i$ corresponding to the label $\mathbf{q}_{i}^{(t)}$.
The coefficients $\{\varphi_{i}\}$ are non-negative weighting hyperparameters,
which enable balancing the loss in recovering the interference terms
and that in recovering the soft estimate of the symbol of interest
$x_{k}$. \textcolor{black}{In particular, for $\varphi_{k}=1$ and $\varphi_{i}=0$
for $i\neq k$, the loss accounts only for the recovery of the symbol
of interest, thus, it is henceforth termed as the \textit{local loss},
where only the data corresponding to the user of interest, i.e., $\big\{ y_{k}^{(t)},\mathbf{q}_{k}^{(t)}\big\},$
is used for training. Alternatively, for $\varphi_{i}=1$ for every
$i\in\{1,\ldots,k\}$, the resultant loss referred to
as the \textit{combined loss}, equally accounts for the
interference terms and the symbol of interest. Using the \textit{combined loss} obviously requires
training data corresponding to both the user of interest and to the preceding
users, i.e., $\big\{ y_{k}^{(t)},\mathbf{q}_{1}^{(t)},...,\mathbf{q}_{k}^{(t)}\big\}$. As such, the \textit{combined loss} explicitly encourages each DNN block to detect its corresponding symbol, while the \textit{local loss} accounts only for the ability of the final DNN block to detect the user's symbol.}


To update the parameters of SICNet, the stochastic gradient descent
(SGD) optimizer is used based on the loss function \eqref{eq:loss}.
The SGD update rule at the $n$-th iteration is given by 
\begin{equation}
\boldsymbol{\theta}_{n+1}:=\boldsymbol{\theta}_{n}-\eta\nabla\mathcal{L}\left(\boldsymbol{\theta}_{n}\right),\label{eq:sgd}
\end{equation}
where $\eta$ denotes the learning rate and $\nabla\mathcal{L}\left(\cdot\right)$
is the gradient of the loss function evaluated at a randomly sampled
mini-batch of the training data. The loss function in \eqref{eq:loss}
is taken over all the DNNs in the SICNet architecture, allowing us to
jointly update the parameters of all $k$ DNN blocks.

In our numerical study we train SICNet relying on a specific SNR
and then test it at different SNRs. This means that the training overhead
can be reduced, since we do not have to retrain the DNN model for different
SNRs. Moreover, SICNet requires only a small dataset for training
to achieve the desired performance. Details of the training SNR, data
size and other hyperparameters are provided for our simulations in Section~\ref{sec:Numerical-Evaluations}.\footnote{\textcolor{black}{The implementation of our SICNet on Python/Tensorflow can be found at https://github.com/ThienVanLuong/SICNet.}}


\vspace{-0.2cm}

\subsection{Discussion\label{subsec:Discussion}}

\vspace{-0.1cm}
 We next discuss some of the advantages and challenges which arise
from the design of SICNet. \textcolor{black}{Firstly, we note that SICNet is specifically
tailored for detecting superimposed signals in  non-orthogonal DL
communications, given its SIC structure. Consequently, when trained
using data corresponding to the same channel for which it is tested,
SICNet is expected to approach the performance of the model-based SIC
detector, as numerically demonstrated in
Section~\ref{sec:Numerical-Evaluations}. Moreover, our scheme is less
sensitive to CSI uncertainty, since it does not rely on the explicit
formulation of the channel's input-output relationship, but rather
learns it implicitly from data. This allows SICNet to achieve superior
performance over the classical detector, when relying on realistic
imperfect CSI.  In particular, SICNet can be trained without knowing
the channel model or requiring the noise to be additive, which makes
it particularly suitable for non-orthogonal systems, where the channel
model is complex, as it is
commonly the case in the presence of hardware
impairments. Furthermore, in contrast to the
model-based SIC detector, SICNet only requires the users to know the
order of the superposition code, rather than the actual power
allocation coefficients of each user in the network. While this
partial knowledge is exploited by SICNet, we numerically show in
Section~\ref{sec:Numerical-Evaluations}  that it is robust to
perturbations in the superposition code.}

\textcolor{black}{An additional benefit of SICNet, discussed in the following section, follows
from its ability to produce soft estimates in a model-agnostic fashion.
In particular, when the model-based SIC is used for producing soft outputs,
it typically relies on approximations of the distribution of the error
term, as in \eqref{eqn:PDFApprox}, due to the difficulty in explicitly
characterizing its PDF. SICNet, which relies on deep learning to produce
its conditional distribution estimates, does not have to know the model of 
the interference and its estimation error, rather it learns  solely from
data. Consequently, once properly trained, SICNet is capable of implicitly
learning to accurately produce bit-wise LLRs for improving the overall decoding performance when combined with
soft-input FEC decoders, as detailed in Section~\ref{sec:online-training}.}

Several challenges are associated with SICNet in its current form.
Being a data-driven implementation of the SIC algorithm, it recovers
the symbols based on the rank-order dictated by the superposition code.
This implies that changing the coding scheme would require adapting
SICNet. Nonetheless, the change in the order of the users or the introduction of a new user in the DL does not necessarily imply that the architecture of SICNet has to be modified, since SICNet is still able to maintain high-integrity detection with mismatched architecture by retraining it with the \textit{local loss} objective, as we numerically demonstrate in Subsection~\ref{subsec:SERRes}. An additional scenario in which SICNet has to be retrained
is when the underlying statistical model of the channel changes. In
particular, SICNet is designed for stationary channels, where the
same mapping can be reliably applied over multiple time instances,
and the channel conditions remain static during both the training and
testing periods. SICNet can be applied reliably even when trained
using channel conditions and SNRs which are different from those used
for testing, as we will numerically demonstrate in Section \ref{sec:Numerical-Evaluations}.
However, when the channel conditions change considerably over time,
one would eventually have to retrain SICNet to maintain reliable
operation. A compelling technique of online training due
to changes in either the channel conditions or the superposition code is to train from coded transmissions in a self-supervised manner, as
proposed in \cite{Shlezinger2020ViterbiNet}, which we carefully adapt
for SICNet in Section~\ref{sec:online-training}.

	\textcolor{black}{Finally, when the number of users increases, the complexity of the conventional SIC escalates due to the need to carry out more interference cancellation steps. Accordingly, the complexity of SICNet - which is reminiscent of the model-based SIC - also scales with the number of users. In such scenarios, one may have to carefully fine-tune the DNN hyperparameters for achieving the desired performance, and utilize DNNs having a large number of inputs, when cancelling the interference of users having lower power. This task is likely to be feasible even for large non-orthogonal networks, since DNNs are inherently compliant with high-dimensional data. In fact, the amalgamation of the SIC algorithm with DNNs in SICNet may allow it to carry out detection more promptly than model-based techniques due to the fact that DNNs conveniently lend themselves to  parallelization. Therefore, this drawback - which SICNet inherits from the model-based algorithm - is expected to be less severe for a data-driven implementation than for the classical SIC algorithm.}
\vspace{-0.2cm}

\section{SICNet Relying on FEC Decoding\label{sec:online-training}}

\vspace{-0.1cm}

In this section, we integrate SICNet with FEC decoding for coded downlink
non-orthogonal systems. In particular, we first discuss how SICNet can produce
LLRs to be used for FEC decoding in Subsection \ref{subsec:SoftSICNet}.
Then, in Subsection \ref{subsec:OnlineFEC} we design an FEC-aided
online training strategy for SICNet in the presence of block fading,
where the proposed FEC-coded receiver can adapt to the variations
of block fading channels without requiring dedicated pilot transmissions.

\vspace{-0.1cm}

\subsection{SICNet with Soft-Decoding}

\label{subsec:SoftSICNet} For coded non-orthogonal DL, the message intended
for user $k$, denoted by the bit vector $\mathbf{b}_{k}$, $k=1,...,K,$
is encoded by a FEC encoder at the transmitter before being mapped
into $M$-ary symbols $x_{k}$. These symbols are then superimposed
onto those of other users via \eqref{eq:x_BS}. It is straightforward
to employ hard FEC decoding to both SIC and SICNet, where information bits,
which are estimated from decoded $M$-ary symbols, are fed directly
to a hard FEC decoder for decoding $\mathbf{b}_{k}$. Note that these
two detectors decode $M$-ary symbols based on hard decisions, as shown
in \eqref{eq:sic_user_k} for SIC and in \eqref{eq:x_hat} for SICNet. Therefore, we now focus on soft decoding relying on
bit-wise LLRs. Moreover, akin to the soft decoder of the classical SIC presented in Subsection \ref{subsec:Successive-Inteference-Cancellat}, we assume that user $k$ only knows his/her coding
scheme, but does not know the codes used by other users. Thus we
allow each user to decode only his/her corresponding message.

As noted in Subsection~\ref{subsec:Successive-Inteference-Cancellat}, using the model-based SIC to produce
soft outputs often results in an inaccurate estimate of the LLRs,
since the errors in recovering the preceding symbols are not accounted
for in the postulated PDF \eqref{eqn:PDFApprox}. As a result, the
coded performance of SIC relying on soft decoding degrades significantly,
as analyzed in Subsection \ref{subsec:CodedBERRes}. To address this
fundamental issue, we propose a soft decoder for SICNet, which directly
computes the LLR of each message bit $\{\beta_{n}\}$ based on the
soft output vector $\boldsymbol{\mathbf{p}}_{k}$ produced by SICNet.
In particular, the fact that SICNet produces $\boldsymbol{\mathbf{p}}_{k}$
given in \eqref{eq:p_i}, whose entries are conditional distribution
estimates, allows the LLRs in \eqref{eq:LLR_bits_SIC} to be computed
via: 
\begin{equation}
L_{n}=\text{log}\frac{\text{Pr}\left(\beta_{n}=0|y_{k}\right)}{\text{Pr}\left(\beta_{n}=1|y_{k}\right)}\approx\log\frac{\sum_{\alpha_{j}\in\mathcal{S}_{n}^{(0)}}p_{k,j}}{\sum_{\alpha_{j}\in\mathcal{S}_{n}^{(1)}}p_{k,j}},\label{eq:LLR_bits}
\end{equation}
where $p_{k,j}=\hat{p}\left(x_{k}=\alpha_{j}|y_{k},\mathbf{p}_{1},\ldots,\boldsymbol{\mathbf{p}}_{k-1}\right)$
is the $j$-th entry of $\boldsymbol{\mathbf{p}}_{k}$
and $j=1,...,M$. The LLRs are then fed to a soft FEC decoder for
decoding $\mathbf{b}_{k}.$

Using SICNet for computing the LLRs builds upon
the ability of DNNs to learn conditional distributions in a model-agnostic
manner from data. Interestingly, this simple extension allows SICNet
to provide higher-accuracy LLRs than the soft decoder of SIC,
which is based on the inaccurate Gaussian approximation, leading to
better coded performance, as demonstrated in Subsection \ref{subsec:CodedBERRes}.

\vspace{-0.1cm}

\subsection{FEC-Aided Online Training}

\label{subsec:OnlineFEC} The combination of SICNet with coded communications
can be exploited to learn to adapt to block fading channel conditions
without requiring dedicated pilot transmissions. Here, we follow the
guidelines proposed in \cite{Shlezinger2020ViterbiNet} to enable
online training of SICNet from decoded codewords in a self-supervised
manner. This strategy exploits the capability of FEC codes to correct detection errors and to provide feedback on the
accuracy of the outputs of SICNet. 

In block fading channels, the channel input-output distribution \eqref{eq:cond_pdf}
remains unchanged within a transmission block, while varying from
one block to another. We assume that those variations are gradual,
i.e. that while the channel can change dramatically over multiple
blocks, the variations between consecutive blocks are limited in the
sense that a symbol detector applicable for one channel block is also
expected to operate adequately under the statistical model of the following
block. Our goal is to allow SICNet adapt to the changes of block fading
over time, where FEC codes are exploited for recovering data labels
used for retraining of SICNet online.

\begin{figure}[tb]
\begin{centering}
\includegraphics[width=\figWidth]{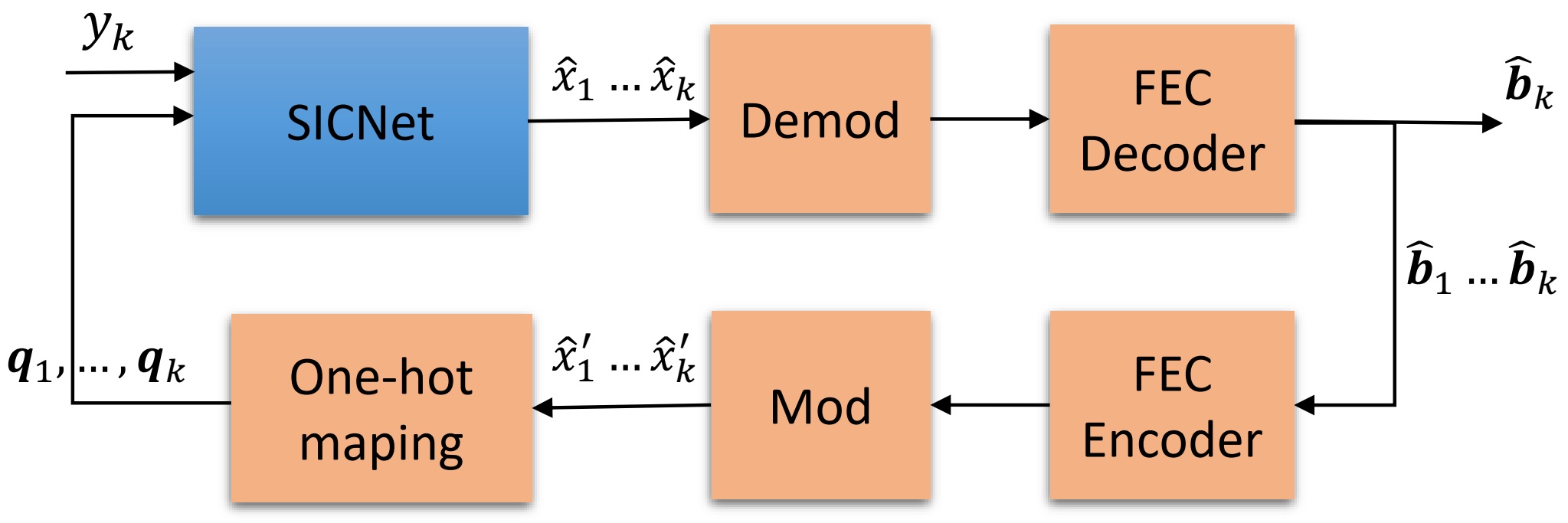} 
\par\end{centering}
\caption{FEC-aided online training model of SICNet. \label{fig:OnlineSICNet}}
\end{figure}

In coded non-orthogonal DL operating over block fading channels,
each fading block corresponds to the transmission of a superimposed
message. In particular, for every fading block, the data bit vectors
$\mathbf{b}_{k}$, $k=1,...,K$ are encoded by an FEC encoder prior
to being modulated into $M$-ary symbols $x_{k}$, which are then
superimposed for transmissions. To characterize the ability of SICNet
to adapt to the block-wise variations in the underlying statistical
model, consider the $k$-th user employing SICNet for symbol
detection, as illustrated in Fig. \ref{fig:OnlineSICNet}. For simplicity,
we focus on the usage of hard estimates. Here, the symbols of the
$k$ users $\hat{x}_{1},...,\hat{x}_{k}$ estimated from the soft
outputs $\mathbf{p}_{1},...,\mathbf{p}_{k}$ of SICNet are demodulated
into uncoded bits, which are then decoded by a FEC decoder to obtain
$k$ estimates of the users' information bits, denoted by $\hat{\mathbf{b}}_{1},...,\mathbf{\hat{b}}_{k}$.
The FEC decoding procedure implies that successful decoding is achievable,
i.e. that $\{\hat{\mathbf{b}}_{i}\}$ are equal to the transmitted
messages, even when $\{\hat{x}_{i}\}$ are different from the transmitted
$\{x_{i}\}$. This property can be exploited for generating the postulated
transmitted symbols as proposed in \cite{Shlezinger2020ViterbiNet},
which can be used to train SICNet. Specifically, the estimated bits
are re-encoded and re-modulated to obtain $M$-ary symbols $\hat{x}_{1}',...,\hat{x}_{k}'$,
which represent the postulated transmitted symbols. As a result, by
mapping $\hat{x}_{1}',...,\hat{x}_{k}'$ into one-hot vectors, we
can generate online labels $\mathbf{q}_{1},...,\mathbf{q}_{k}$ corresponding
to the current channel output $y_{k}$ for retraining SICNet without
requiring dedicated pilot transmissions, making it adaptable to the
variations of block fading channels.

In general, the proposed online training mechanism requires user $k$
to know the channel coding schemes of its preceding users indexed
by $1,...,k-1$, as it must decode their corresponding messages in
order to provide the labels required for evaluating the loss function \eqref{eq:loss}.
When this knowledge is not available, e.g., as in the scenario discussed
in Subsection~\ref{subsec:SoftSICNet}, the $k$-th user can still
retrain SICNet using only his/her own decoded message by setting the
loss measure not to account for the recovery of the interference.
This can be done by setting $\varphi_{i}=0$ for $i\neq k$ in \eqref{eq:loss},
i.e., the \textit{local loss} is used. Furthermore, while the proposed
online training mechanism is detailed for SICNet using hard FEC decoders,
it can also be applied when SICNet is combined with soft FEC decoders.
In that case, one should simply replace the demodulation block of Fig. \ref{fig:OnlineSICNet} with 
the LLR calculation block. Here, the
soft outputs of SICNet, namely, $\mathbf{p}_{1},...,\mathbf{p}_{k}$,
are used for computing the LLRs of $k$ users as presented in the
previous subsection. The proposed online training SICNet based on
both hard and soft FEC decoding can track the variations of block
fading without any CSI estimation, whilst this cannot be achieved by the conventional
SIC, as  numerically demonstrated in Subsection \ref{subsec:CodedBERRes}.

Finally, we note that the proposed online training scheme builds upon
successful FEC decoding following \cite{Shlezinger2020ViterbiNet}.
Nonetheless, SICNet can also be combined with alternative techniques to allow a DNN-aided receiver to track time-varying
channel conditions at a modest overhead. These include the application
of meta-learning for optimizing the hyperparameters of the training algorithm
\cite{park2020meta}; the pre-training of multiple receivers as a
deep ensemble \cite{raviv2020data}; and the usage of soft symbol-level
outputs, rather than FEC decoding, as a measure of confidence for producing
labels from data, as proposed in \cite{Sun2021Generative,finish2022symbol}. We leave the
study of the combination of SICNet with these methods to facilitate
online training for future investigations.

\vspace{-0.2cm}

\section{Numerical Evaluations\label{sec:Numerical-Evaluations}}

\vspace{-0.1cm}
 In this section, we numerically evaluate the performance of SICNet,
comparing it to the model-based SIC algorithm. Both perfect and
imperfect CSI are considered. In addition to a linear Gaussian channel,
we also consider a non-linear quantized Gaussian channel. In the following,
we introduce the parameters used for evaluating SICNet, followed
by its symbol error rate (SER)
when used for symbol detection, as well as the coded bit error rate (BER),
when combined with FEC decoding. 
\vspace{-0.2cm}

\subsection{Implementation Setting}

\subsubsection{Simulation Parameters}

\begin{table}
\caption{A summary of simulation parameters\label{tab:para}}

\centering{}%
\begin{tabular}{|l|r|}
\hline 
Parameter  & Value\tabularnewline
\hline 
\hline 
SICNet for non-orthogonal DL user $k=K$  & 3\tabularnewline
\hline 
Modulation order $M$  & 2\tabularnewline
\hline 
Power coefficients of three users $P_{1}$-$P_{2}$-$P_{3}$  & 16-4-1\tabularnewline
\hline 
Hidden nodes of DNN block 1  & 24-12\tabularnewline
\hline 
Hidden nodes of DNN block 2  & 32-16\tabularnewline
\hline 
Hidden nodes of DNN block 3  & 48-32\tabularnewline
\hline 
Activation function for hidden layers  & \textcolor{black}{ReLU \cite{DeepIM2019}}\tabularnewline
\hline 
Activation function for output layers  & \textcolor{black}{Softmax \cite{Luong2020MC-AE}}\tabularnewline
\hline 
Training SNR $1/\sigma^{2}$  & 6 dB\tabularnewline
\hline 
Learning rate $\eta$  & 0.001\tabularnewline
\hline 
Batch size  & 100\tabularnewline
\hline 
Number of training epochs  & 200\tabularnewline
\hline 
Training data size  & 5000\tabularnewline
\hline 
Testing data size  & $10^{6}$\tabularnewline
\hline 
Optimizer  & Adam \cite{Kingma2014AdamAM}\tabularnewline
\hline 
\end{tabular}
\end{table}

The parameters used in our simulations of SICNet are summarized in
Table~\ref{tab:para}. We consider a non-orthogonal DL system supporting $K=3$
users, focusing on user $k=3$, which involves the highest number of
interference cancellation steps. The BS sends BPSK symbols to all
users, i.e., $M=2$. The power coefficients for user 1, 2, and 3 are
$P_{1}=16$, $P_{2}=4$, and $P_{3}=1$, respectively. The power
coefficients remain unchanged during the training and testing
phases \textcolor{black}{unless otherwise stated.} Each DNN block of
SICNet is comprised of two fully-connected hidden layers as
illustrated in Fig.~\ref{fig:DNN_arch1}, whose dimensions are provided
in Table~\ref{tab:para}. SICNet is trained in an end-to-end fashion
using the Adam optimizer \cite{Kingma2014AdamAM} with a learning rate
of $\eta=0.001$. The training set is comprised of as few as 5000
symbols generated from a channel at an SNR of 6 dB, which was
empirically shown to offer a good performance, when testing over
channels having various SNR levels\footnote{\textcolor{black}{Here, we
set the channel to satisfy $\mathbb{E}\big[|h_{k}|^{2}\big]=1$. As a
result, the SNR is defined for the user of interest
(i.e., user $k=3$) as $1/\sigma^{2}$, where $\sigma^{2}$ is the
variance of the Gaussian noise.}}. Both loss types presented in
Subsection
\ref{subsec:Training-SICNet}, namely, \textit{local loss} and \textit{combined
loss}, are considered. 
Finally, the remaining hyperparameters, such as, the testing data size,
epochs, and batch size, are detailed in Table~\ref{tab:para}. \textcolor{black}{The hyperparameters in Table~\ref{tab:para} have been selected using the grid-search method in order to provide the best performance, while minimizing complexity and training time. For example, we have tentatively trained our SICNet at different training SNRs, namely $3, 4, \ldots, 10$ dB, and found that $6$ dB provides the best BER performance in a range of testing SNRs of interest.} 

\subsubsection{Channel Models}

We consider two channel models: a linear Gaussian channel as in \eqref{eq:y_k},
and a non-linear quantized Gaussian channel.  For both channels, we assume that the channel coefficient of
user 3 is static by simply setting $h_{3}=1$ over both the training and
testing phases. As such, for the linear Gaussian channel, the received
signal of user $3$ is written as $y_{3}=x+w_{3}$, where $x=\sum_{i=1}^{3}\sqrt{P_{i}}x_{i}$,
while for the quantized Gaussian channel, it is given by $y_{3}=\mathcal{Q}\left(x+w_{3}\right)$ \cite{jeon2018supervised},
where $\mathcal{Q}(\cdot)$ represents a 3-bit quantization, given
by 
\begin{equation}
\mathcal{Q}\left(u\right)=\begin{cases}
\text{sign}\left(u\right) & \left|u\right|<2,\\
\text{3\ensuremath{\times}sign}\left(u\right) & 2<\left|u\right|<4,\\
\text{5\ensuremath{\times}sign}\left(u\right) & 4<\left|u\right|<6,\\
\text{7\ensuremath{\times}sign}\left(u\right) & \left|u\right|>6.
\end{cases}\label{eq:Q_quantize}
\end{equation}

To model different levels of CSI, the channel known to the model-based
receiver, and used to generate training for SICNet, is given by a
noisy estimate $\hat{h}_{3}=h_{3}+e$, where $h_{3}=1$ is the actual
channel used during testing, and $e\sim\mathcal{N}\left(0,\epsilon^{2}\right)$
is the channel estimation error. In particular, the conventional SIC
uses $\hat{h}_{3}$ for decoding instead of $h_{3}$, while SICNet
is trained using samples generated from the erroneous channel, with
$\hat{y}_{3}=\hat{h}_{3}x+w_{3}$ or $\hat{y}_{3}=\mathcal{Q}\big(\hat{h}_{3}x+w_{3}\big)$
for linear and quantized Gaussian channels, respectively. We set $\epsilon^{2}=0$
for {\em perfect CSI}, and $\epsilon^{2}=0.01$ for {\em imperfect
CSI}. 

\textcolor{black}{In order to obtain the data used for training SICNet, we first randomly generate the data symbols sent to the $K$ users, $\{x_{k}\}_{k=1}^{K},$ which are then used to obtain the superimposed code $x$ based on \eqref{eq:x_BS}. Through the channel of user $k$, we obtain the corresponding received signal $y_k$, which is combined with the symbols sent from the BS to users to create the training data set, here for user $k$.}

\vspace{-0.2cm}

\subsection{SER Performance}

\label{subsec:SERRes} \vspace{-0.1cm}

\begin{figure}[tb]
\begin{centering}
\includegraphics[width=\figWidth]{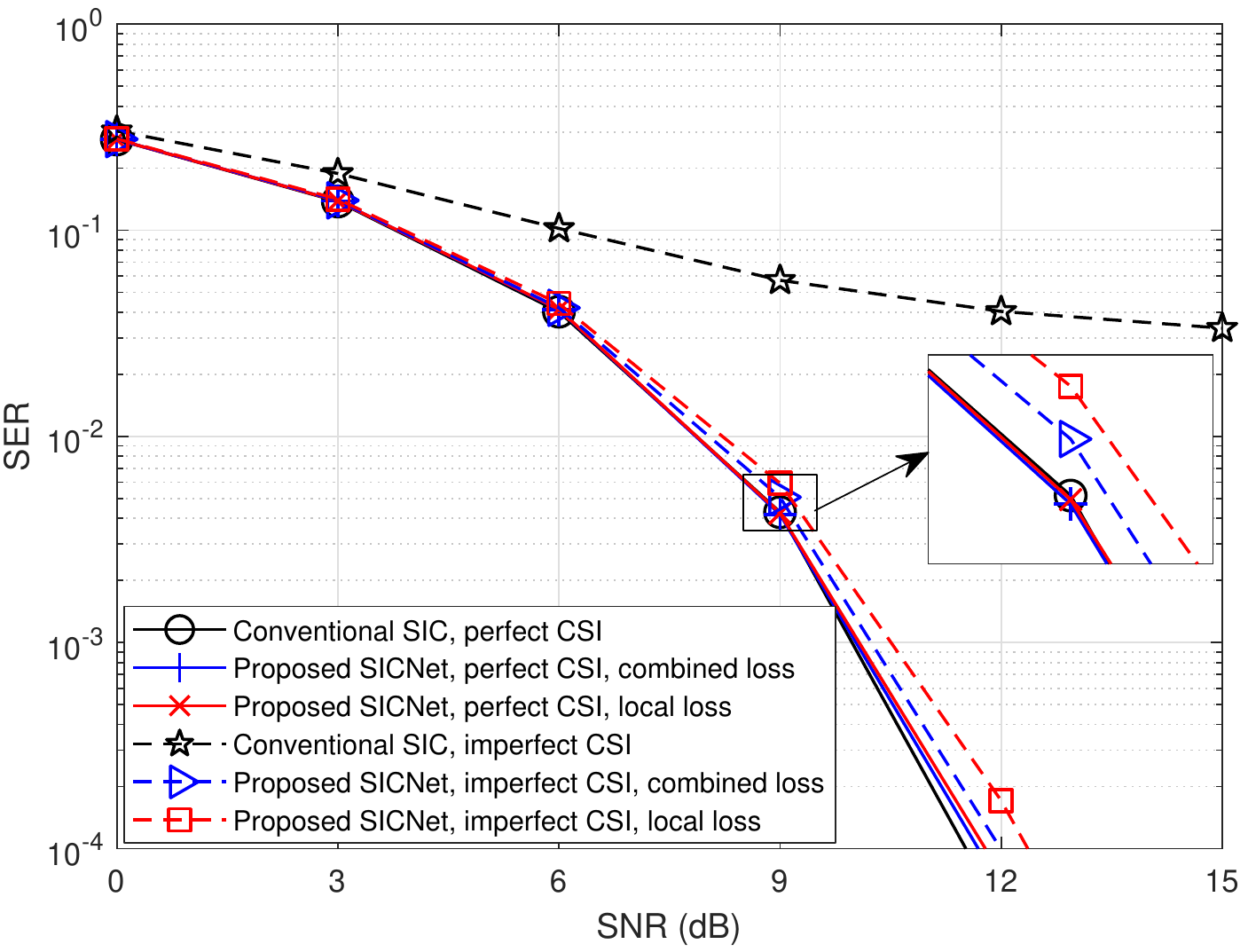} 
\par\end{centering}
\caption{SER comparison between the proposed SICNet and conventional SIC under
linear Gaussian channels with both perfect and imperfect CSI. Here,
the proposed SICNet is trained with both \textit{local} and \textit{combined} losses.
\label{fig:imperfect_csi}}
\end{figure}

We first numerically evaluate the SER of SICNet compared to the model-based
SIC, when used for symbol detection, i.e. to produce hard decisions
of the transmitted symbols. Fig. \ref{fig:imperfect_csi} depicts
our SER comparison between the proposed SICNet trained with both loss
measures, and the conventional SIC under the linear Gaussian channel,
for both perfect and imperfect CSI conditions. We observe in Fig.
\ref{fig:imperfect_csi} that when trained and tested for the same
channel, SICNet achieves a SER performance approaching that of the conventional
SIC operating with perfect knowledge of the channel model and its
parameters. This indicates that our DNN-aided detector learns to implement
the model-based SIC algorithm from data, while being trained for a
single SNR level. Furthermore, under CSI uncertainty, the SER of the
conventional SIC is notably degraded, while our SICNet still achieves
accurate detection, where its SER is within a minor gap of its performance
with perfect CSI. For example, at a SER of $10^{-3},$ the channel
imperfection causes an SNR loss of less than 0.5 dB for SICNet compared
to the perfect CSI condition. Finally, despite only using the local
data for training, the \textit{local loss} achieves SER values within
a minor gap to that of the \textit{combined loss} under both CSI scenarios. 

\begin{figure}[tb]
\begin{centering}
\includegraphics[width=\figWidth]{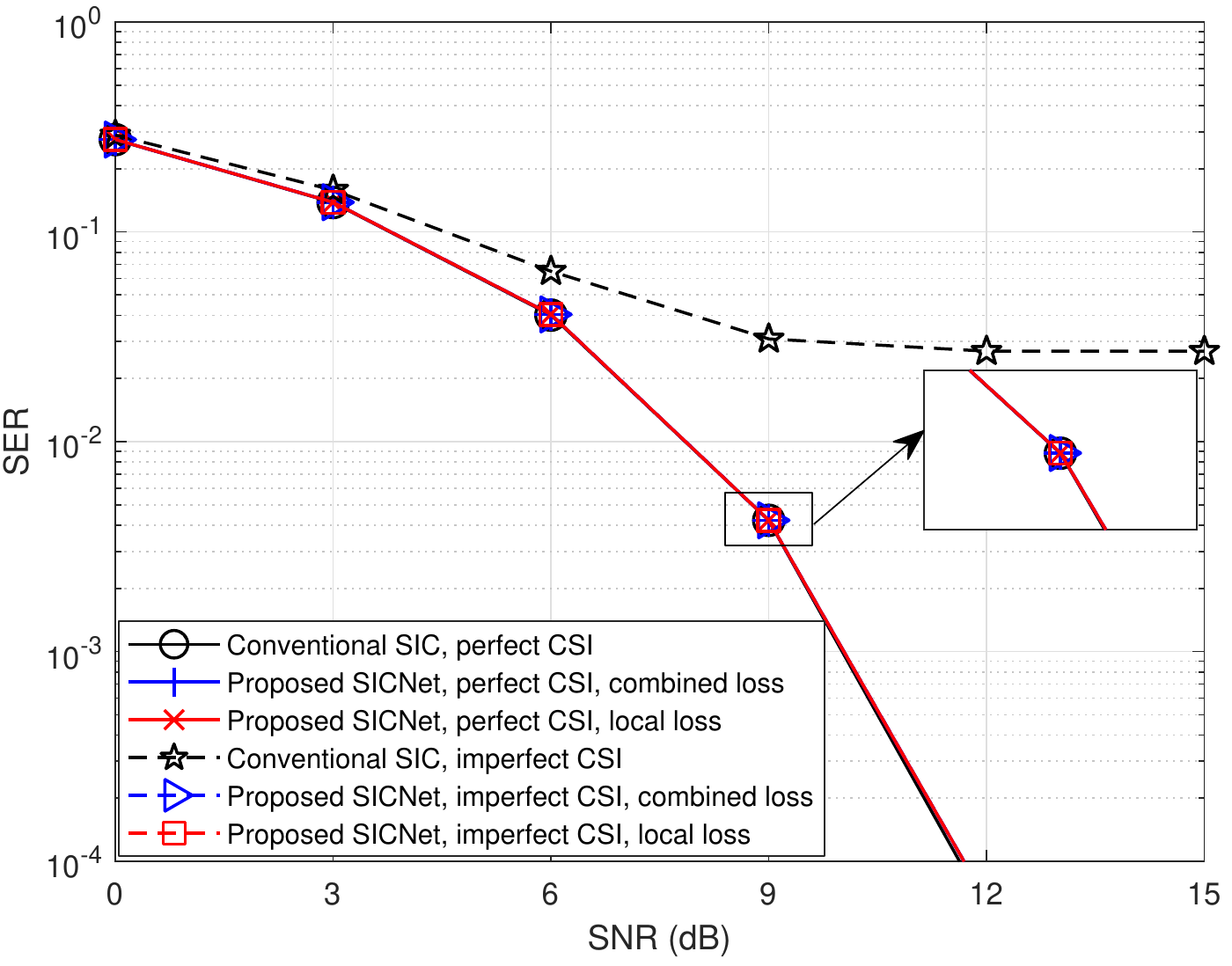} 
\par\end{centering}
\caption{SER comparison between the proposed SICNet and conventional SIC under
quantized Gaussian channels with both perfect and imperfect CSI. Here,
the proposed SICNet is trained with both \textit{local} and \textit{combined} losses.
\label{fig:quantized}}
\end{figure}

Fig. \ref{fig:quantized} compares the SER of the proposed SICNet
using the two losses with the model-based SIC under the quantized
Gaussian channel, for both perfect and imperfect CSI conditions. Again,
it is observed in Fig. \ref{fig:quantized} that our SICNet achieves
similar SER values as the model-based scheme, when the CSI is perfect.
Although the channel is non-linear, the quantization resolution is
sufficient to allow the interference to be approximately canceled
by subtraction, and thus the model-based SIC algorithm still achieves
accurate recovery here. However, when the CSI is imperfect, our scheme
significantly outperforms its conventional counterpart, which suffers from a relatively
high error floor, i.e., $>10^{-2}$. \textcolor{black}{In particular, SICNet is hardly affected here by the imperfect CSI. This is likely to be due to the fact that the presence of quantization results in the training data under imperfect CSI being quite similar to that generated from the true channel.} This validates the efficiency
of SICNet under quantized Gaussian channels, even with imperfect CSI,
where the model-based SIC achieves poor SER performance. These observed
gains follow from the usage of DNNs for decoding soft-information
for each symbol, which allows SICNet to learn to implement SIC without
relying on channel modelling. Again, it is observed via Fig. \ref{fig:quantized}
that using the \textit{local loss}, the proposed SICNet achieves SER
values, which are similar to the \textit{combined loss} under quantized
Gaussian channels.

\begin{figure}
\begin{centering}
\includegraphics[width=\figWidth]{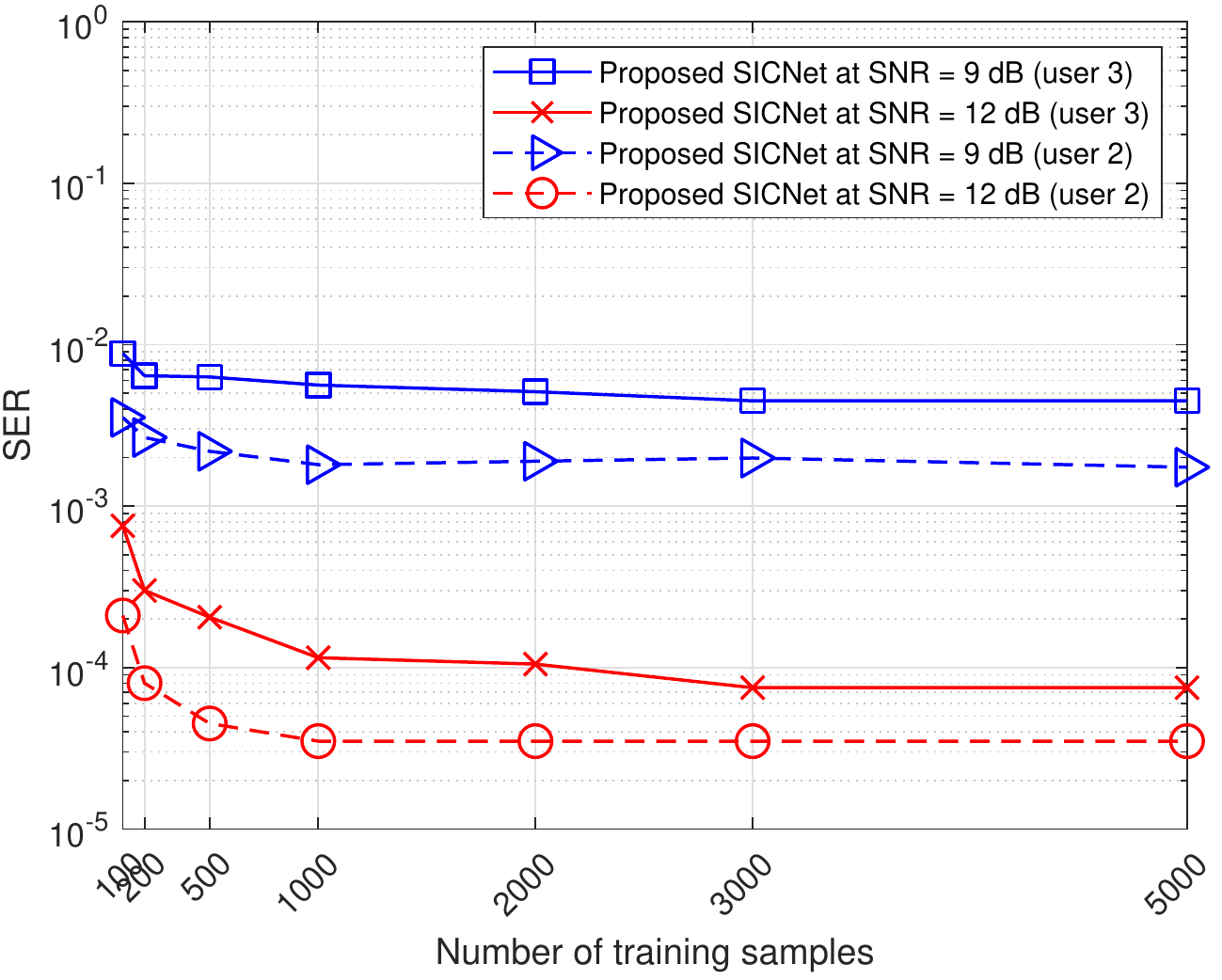}
\par\end{centering}
\caption{\textcolor{black}{SER performance of the proposed SICNet for user 2 and user 3, trained using different number of training samples, under linear Gaussian channels. Here, SICNet is trained with the aid of the \textit{local} loss using data from the true underlying channel.}
\label{fig:ser_datasize}}
\end{figure}

In the numerical results presented in
Figs.~\ref{fig:imperfect_csi}-\ref{fig:quantized}, SICNet is trained
using merely $5000$ labeled samples, representing, e.g. pilots and
preamble sequences routinely used in wireless  schemes. \textcolor{black}{In order
to numerically quantify the number of samples required for training
SICNet corresponding to different users having different order in the superposition code, we next compare the training of user 3 to that of user 2. We depict in Fig.~\ref{fig:ser_datasize} the accuracy of SICNet when
trained using different numbers of training samples at the SNRs of 9 dB
and 12 dB. It is observed from Fig.~\ref{fig:ser_datasize} that for both users, SICNet
can in fact be accurately trained with much less than $5000$
samples, and often as few as $1000$ samples are sufficient. For example, at the SNRs of 9dB and 12 dB, SICNet
requires only $200$ samples and $1000$ samples, respectively, for
achieving a SER performance which is very close to that trained using
$5000$ samples. This is due to the fact that SICNet is a hybrid
model-based and data-driven scheme, which incorporates the SIC
structure into its design, allowing us to significantly reduce the
amount of training, which can be translated into using less pilots and hence 
improved spectral efficiency. Furthermore, the results reported in Fig.~\ref{fig:ser_datasize} indicate that although different users may require different numbers of DNN blocks, using a relatively small amount of pilots for training SICNet is sufficient for different users achieving their desired performance.}

\begin{figure}[tb]
\begin{centering}
\includegraphics[width=\figWidth]{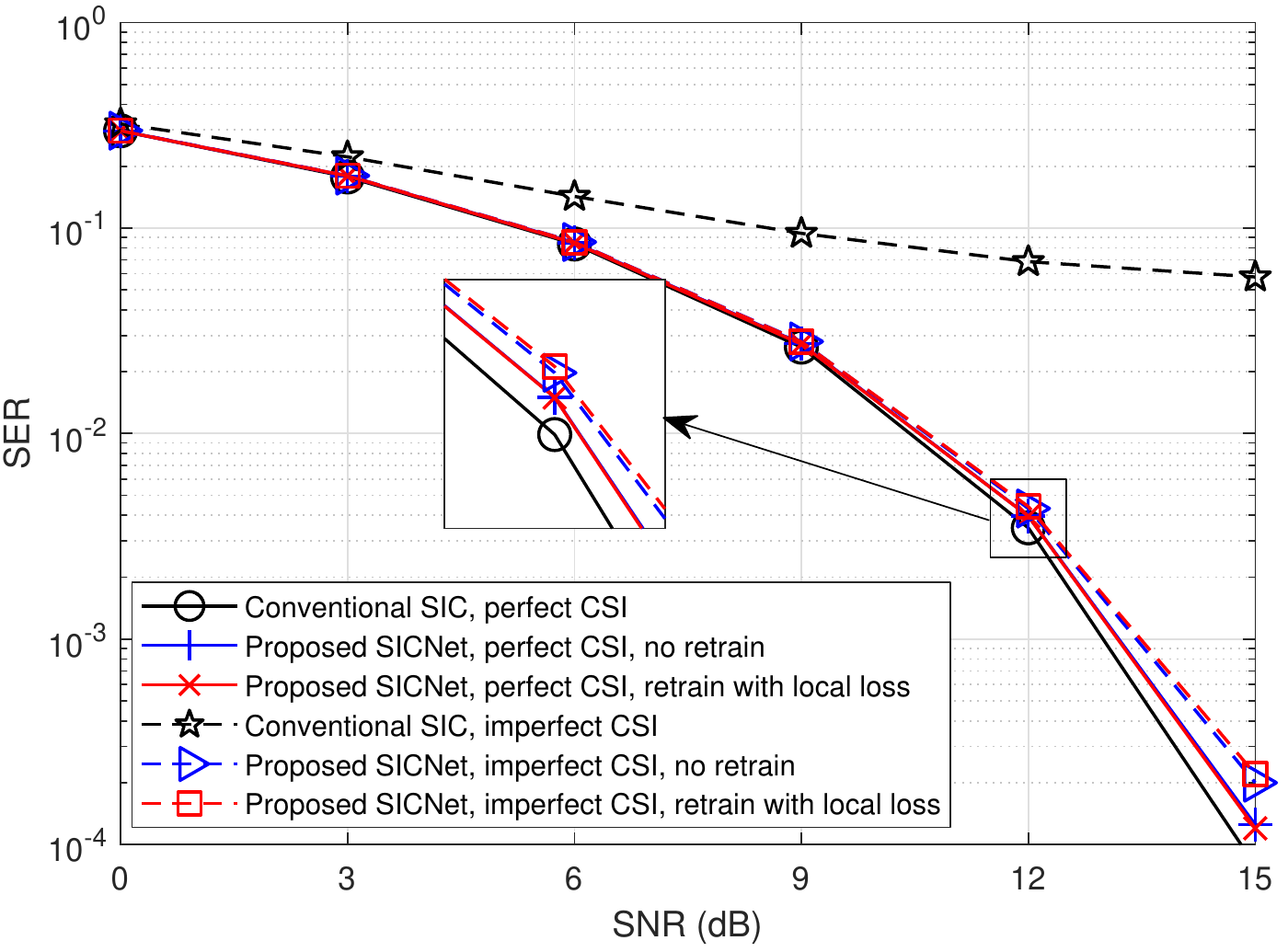} 
\par\end{centering}
\caption{SER performance of the proposed SICNet of user 3 when adding the new user 4 to the existing non-orthogonal DL system of three users. Herein, linear Gaussian channels with both perfect and imperfect CSI are considered, conventional SIC is included for comparison, and SICNet is trained only with the \textit{local loss}.
\label{fig:add_user}}
\end{figure}

\textcolor{black}{We proceed by numerically evaluating the robustness of SICNet to
perturbations in the superposition coding scheme, focusing on linear
Gaussian channels, trained with the {\em local loss} objective. Since
the architecture of SICNet is dictated by the order of the power
assignment among the users, our aim here is to study the ability of
SICNet to handle modifications in this order by retraining. In
Fig.~\ref{fig:add_user} we depict the SER performance of the SICNet of
user 3, i.e. of the user of interest, when adding the new user 4 to
the existing system of three users. In particular, this user has a
power coefficient of $P_4=1/9$, which is lower than that of all the 
existing users, hence resulting in the new order of $P_1>P_2>P_3>P_4$. In this
context, the architecture of SICNet detailed in Table~\ref{tab:para}
still matches the superposition coding, since the power coefficient of
user 3 is still the third lowest in the new system, i.e., the majority
of interference emanates from users 1 and 2. In
Fig.~\ref{fig:add_user}, we investigate the associated SER performance both with and
without retraining using the \textit{local loss} in the new 4-user
system. For retraining, a new dataset of 5000 samples is used that also takes
the impact of the added user into account. Observe in
Fig.~\ref{fig:add_user} that the new source of interference results in
some SER degradation compared to the scenario without this user seen in
Fig.~\ref{fig:imperfect_csi}. However, SICNet succeeds in maintaining an accurate
detection both with and without retraining, while exhibiting improved robustness to imperfect CSI compared to the model-based SIC.}  

\begin{figure}[tb]
\begin{centering}
\includegraphics[width=\figWidth]{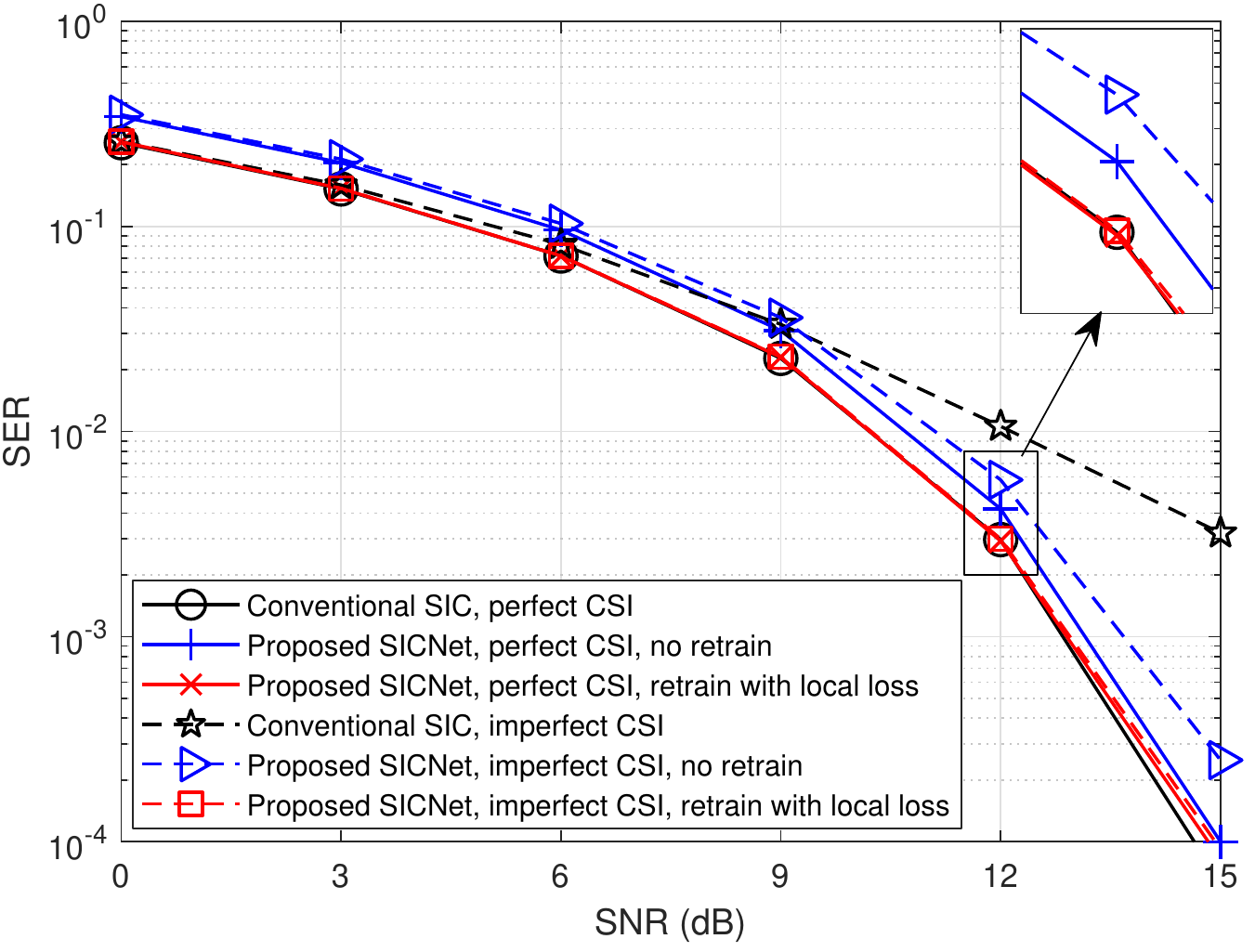} 
\par\end{centering}
\caption{SER performance of the proposed SICNet of user 3 when changing the rank-order of users in the existing non-orthogonal DL system of three users, under linear Gaussian channels with both perfect and imperfect CSI. Here, conventional SIC is also included for comparison, while SICNet is trained only with the \textit{local loss}.
\label{fig:change_order}}
\end{figure}

\textcolor{black}{In the scenario considered in Fig.~\ref{fig:add_user}, the introduction of the new user 4 does not affect the power assignment order of users 1-3, mainly resulting in lightly increased interference treated as additional effective noise. Fig.~\ref{fig:change_order} illustrates the SER performance of  SICNet for user 3, when the order of users changes, i.e., the superposition code used by the BS is modified. In particular, we keep the transmit power of user 2 and 3 unchanged, i.e., $P_2 = 4$ and $P_3=1$, while user 1 now has the lowest power of $P_1=1/9$. As such, the power order of the users has changed from $P_1> P_2>P_3$ into $P_2>P_3>P_1$. In this simulation, we investigate the SER of user 3, both with and without retraining, when its SICNet architecture remains that detailed in Table~\ref{tab:para}, i.e. it does not match the new downlink systems. In order to make conventional SIC adapt to such a change of the users' power allocation in this scenario, user 3 has to know its new power order, which is now $P_2>P_3>P_1$. Then, it would detect and cancel the symbol of user 2 first before detecting its own symbol using the ML detection of \eqref{eq:sic_user_k}. As such, compared to the original power order, user 3 does not have to detect the signal of user 1, who previously had the highest power. Similar to the scenario of adding a new user in Fig.~\ref{fig:add_user}, we do not have to change the architecture of SICNet. However, here we observe that retraining with the aid of both perfect and imperfect data using the \textit{local loss} allows SICNet  to continue approaching the performance of the model-based SIC operating with the aid of perfect CSI. In particular, the usage of the \textit{local loss}, which accounts solely for the desired local symbol results in SICNet learning from data to overcome its mismatched interconnection of building blocks, without enforcing its first DNN block to recover the interfering symbol of user 1.}  

\begin{figure}[tb]
\begin{centering}
\includegraphics[width=\figWidth]{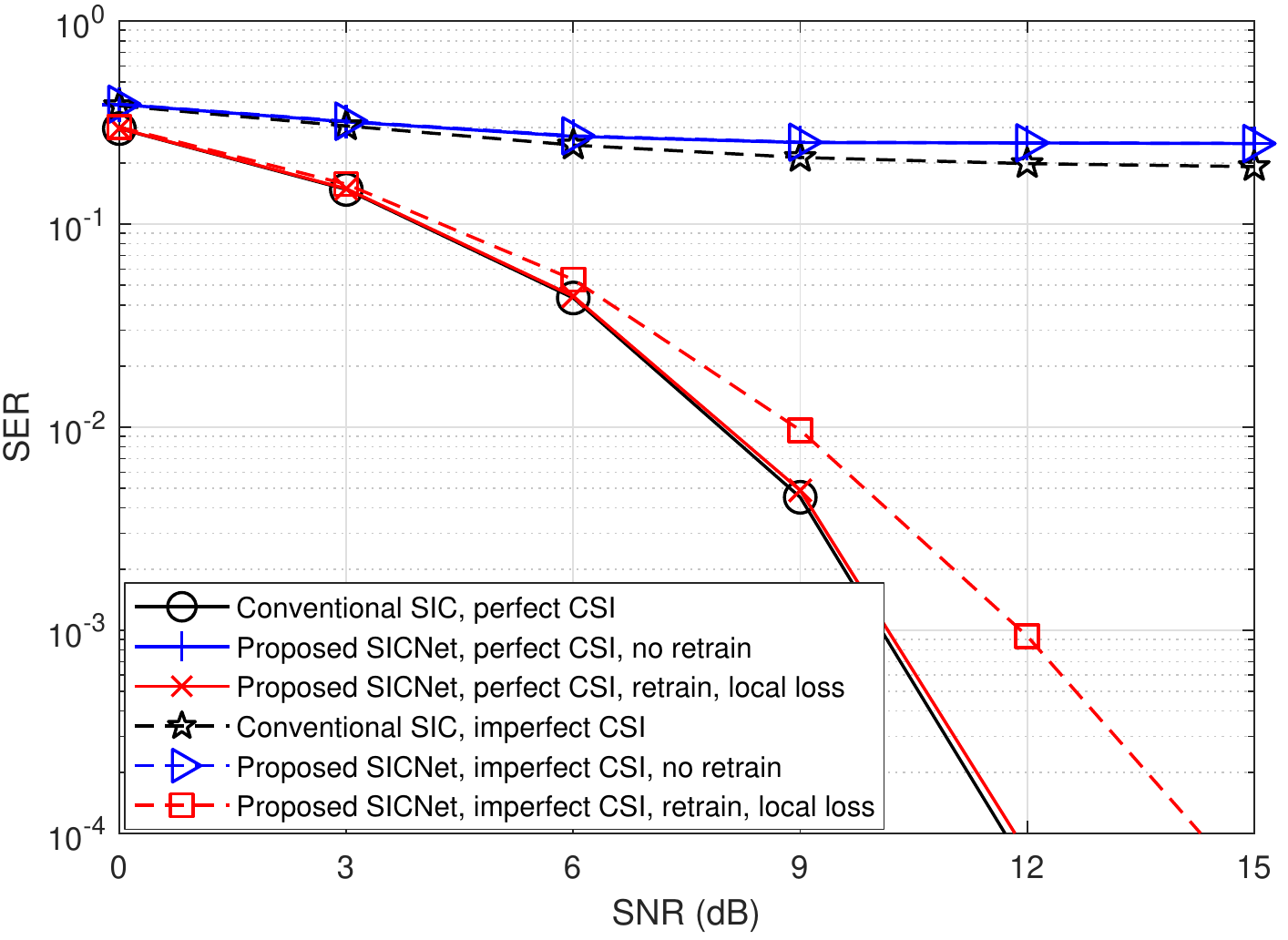} 
\par\end{centering}
\caption{SER performance of the proposed SICNet of user 3 when adding user 4 to the existing system of three users, making the power assignment order changed. Here, conventional SIC is also included for comparison, while SICNet is trained only with the \textit{local loss}. Linear Gaussian channels with both perfect and imperfect CSI are used.
\label{fig:add_user_change_order}}
\end{figure}

Next, we consider the scenario in which the non-orthogonal downlink system changes in both the number of users and their power assignment order. In Fig. \ref{fig:add_user_change_order} we evaluate the case of adding user 4 with $P_4 = 64$, while $P_1$, $P_2$, and $P_3$ remain unchanged, as in Table \ref{tab:para}. Hence, the introduction of user 4 yields a new power assignment order of $P_4>P_1>P_2>P_3$. In contrast to the previous order, $P_3$ is now assigned the fourth lowest power, i.e., it  has  interference from three users. The straight-forward application of SICNet here is  to rebuild its structure by adding one more DNN block. However, as we are focused here on the robustness of SICNet to modifications in the downlink setup,  we keep the existing structure of SICNet with three blocks, detailed in Table \ref{tab:para}.  By observing Fig.~\ref{fig:add_user_change_order}, we note that  SICNet trained for the original downlink setup with three users no longer reliably detects the desired symbols. However, the same SICNet architecture can still approach the accuracy of the model-based SIC with perfect CSI by retraining it with data corresponding to the new DL configuration. These results demonstrate that while the architecture of SICNet is determined by the superposition code, it is robust to modifications in the code and the network setup, and can be utilized for different power assignments by retraining.

\begin{figure}[tb]
\begin{centering}
\includegraphics[width=\figWidth]{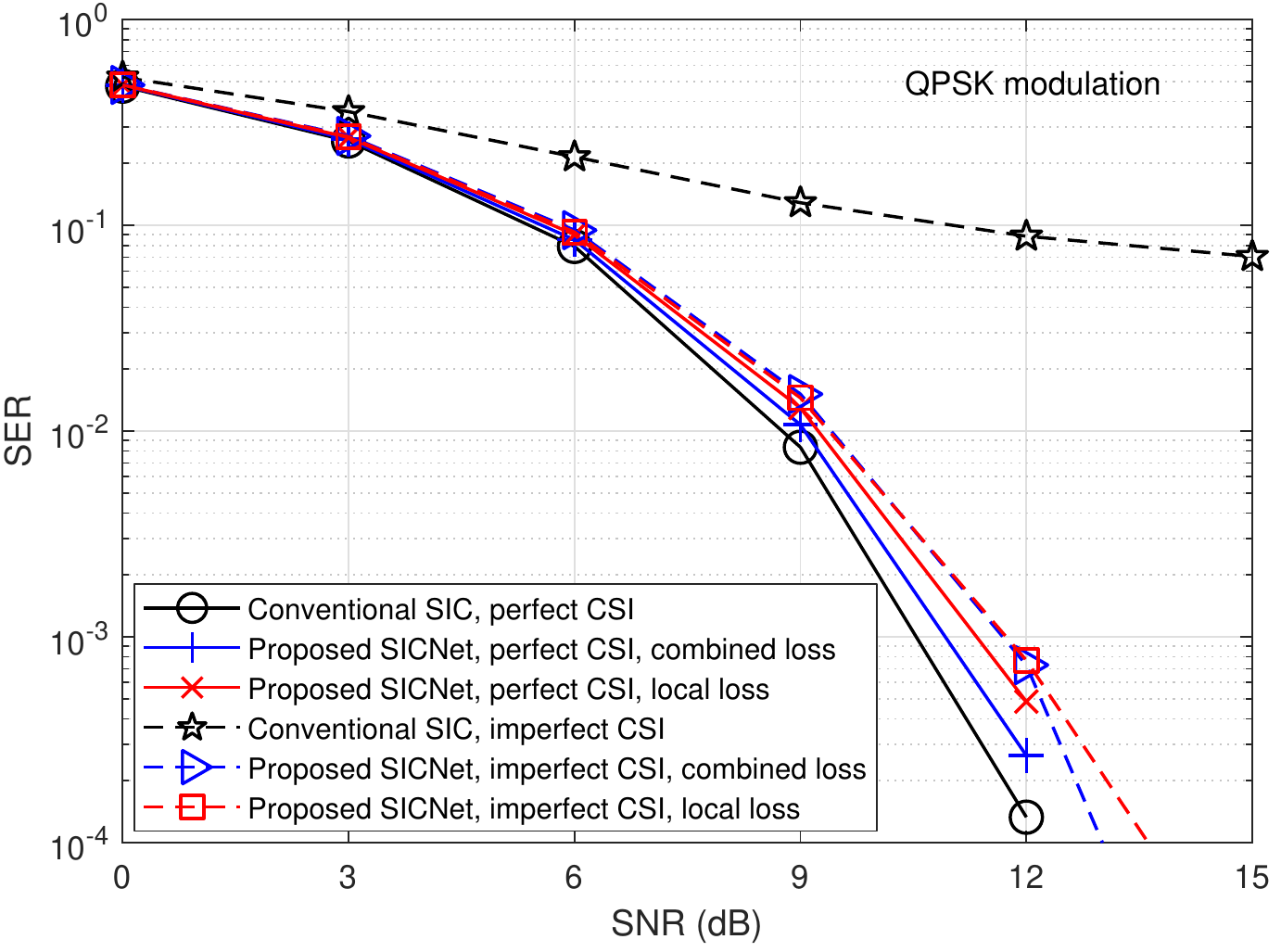} 
\par\end{centering}
\caption{\textcolor{black}{SER comparison between the proposed SICNet and conventional SIC using QPSK modulation, under
linear Gaussian channels with both perfect and imperfect CSI. Here,
the proposed SICNet is trained using both the \textit{local}
and \textit{combined} losses.}
\label{fig:ser_qpsk}}
\end{figure}

 \textcolor{black}{Finally, we demonstrate that while the preceeding
 numerical evaluations focus on real-valued BPSK symbols, SICNet can
 be applied to arbitrary complex modulation schemes. To that aim, in
 Fig.~\ref{fig:ser_qpsk}, we investigate the SER performance of SICNet
 when detecting complex-valued $M$-ary symbols (QPSK modulation), in
 comparison to the conventional SIC detector under linear Gaussian
 channels. Here, a static complex channel $h_3=0.4472 + 0.8944j$ is
 employed instead of the unitary channel used in the aforementioned
 BPSK simulations. Moreover, in contrast to the BPSK case, the DNN
 blocks are now fed with the real and imaginary components of the
 complex received signal $y_3$. The training parameters of
 Table~\ref{tab:para} are reused, except for the training SNR and the
 number of training epochs, which are now 8 dB and 250 epochs in this
 simulation. Both perfect and imperfect CSI scenarios are
 considered. Our SICNet is trained using both the combined and local
 loss. We observe in Fig.~\ref{fig:ser_qpsk} that our scheme applied
 to complex-valued symbols still performs well under both perfect and
 imperfect CSI conditions. In particular, similar to the BPSK results
 of Fig.~\ref{fig:imperfect_csi}, SICNet approaches the model-based
 SIC under perfect CSI, while outperforming this baseline under
 imperfect CSI. Additionally, the performance of our scheme trained
 using the \textit{local} loss is close to that trained employing
 the \textit{combined} loss. These observations confirm that the
 proposed SICNet is also efficient for complex-valued modulated
 symbols.}

\vspace{-0.2cm}

\subsection{Coded BER Performance}

\label{subsec:CodedBERRes} \vspace{-0.1cm}
We numerically evaluate SICNet in a coded communications scenario,
where its outputs are used by a FEC decoder to recover the transmitted
bits. Here, we consider only a linear Gaussian channel, for which
the model-based SIC algorithm is designed. We employ a 1/2-rate convolutional
code using the octally represented generator polynomials $[7;5]$, while utilizing both
hard and soft FEC decoders. We also assume that both SICNet and classical
SIC are unaware of the coding schemes of the preceding users, i.e., the
FEC decoder is used to decode the data bits of the user of interest only.
Accordingly, we train SICNet with the \textit{local loss} measure,
which was shown in the previous subsection to achieve similar performance
to that of training with the \textit{combined loss}, without requiring
any knowledge of the coding schemes of the other users sharing the channel
resources.

\begin{figure}[tb]
\begin{centering}
\includegraphics[width=\figWidth]{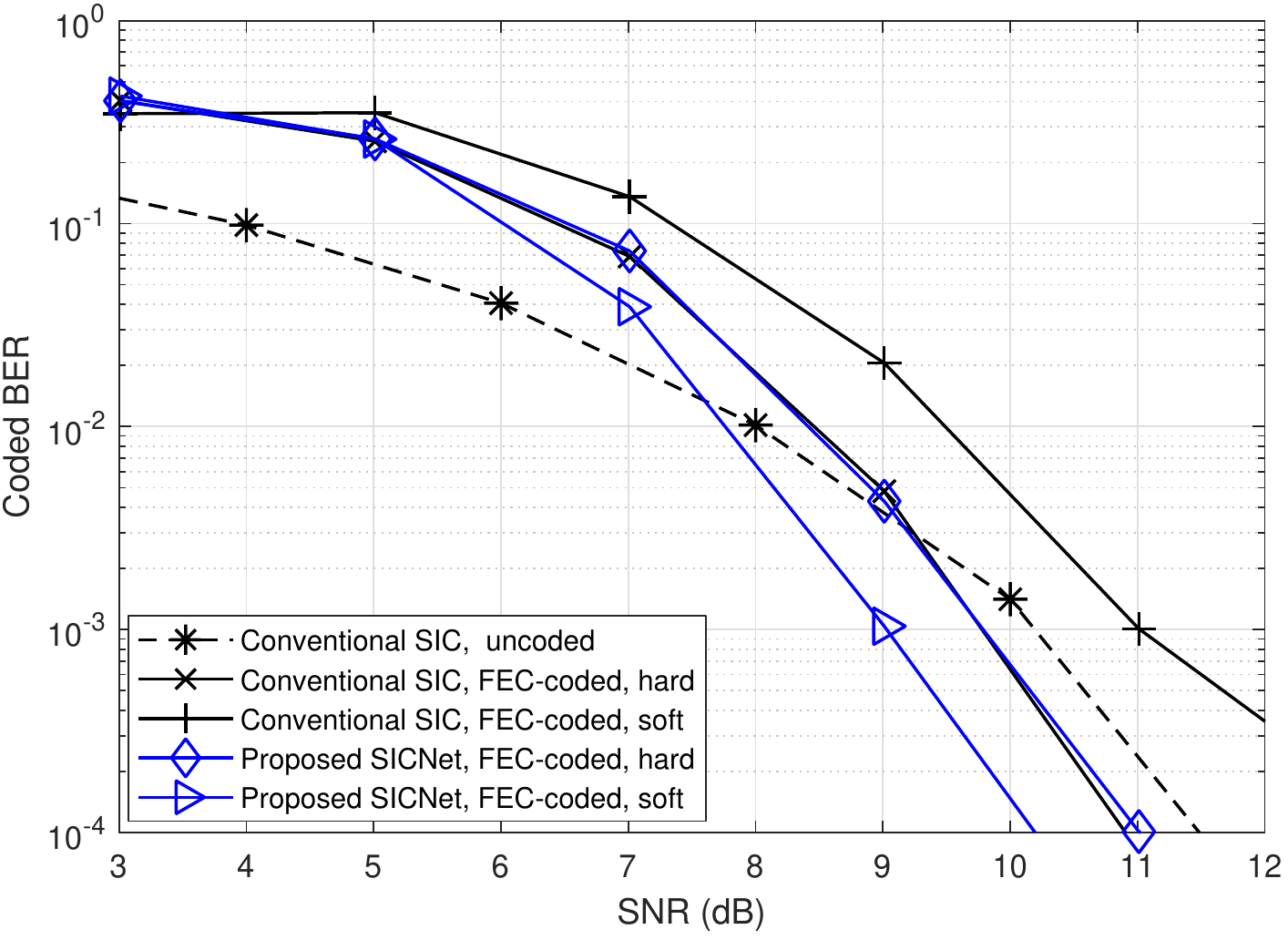} 
\par\end{centering}
\caption{Coded BER comparison between proposed SICNet and conventional SIC
using the 1/2-rate convolutional code under linear Gaussian channels.
Here, the proposed SICNet detector is trained with the \textit{local
loss}. \label{fig:cc_coded}}
\end{figure}

Fig. \ref{fig:cc_coded} compares the coded BER performance of our
proposed SICNet and that of the classical SIC. As user $3$ only knows
his/her coding scheme, he/she does not decode the interference, i.e., the
FEC decoder is used for decoding his/her own data bits only in both SICNet
and its model-based counterpart. It is observed via Fig. \ref{fig:cc_coded}
that the power of FEC coding allows SICNet to achieve improved
accuracy at sufficiently high SNRs over the uncoded scheme, where
the soft decoder achieves a better BER than the hard decoder. Moreover,
using a soft decoder, the proposed FEC-coded SICNet outperforms the
conventional counterpart both for hard and soft decoders, while both
schemes exhibit a similar BER, when a hard-decoder is used. 
These numerical observations demonstrate the ability of SICNet, which
operates in a model-agnostic manner, while learning its mapping from
data, to produce bit-wise LLRs of higher accuracy compared to those
computed by the model-based SIC method, that relies on an approximation
of the distribution of the decontaminated channel output.


\begin{figure}[tb]
\begin{centering}
\includegraphics[width=\figWidth]{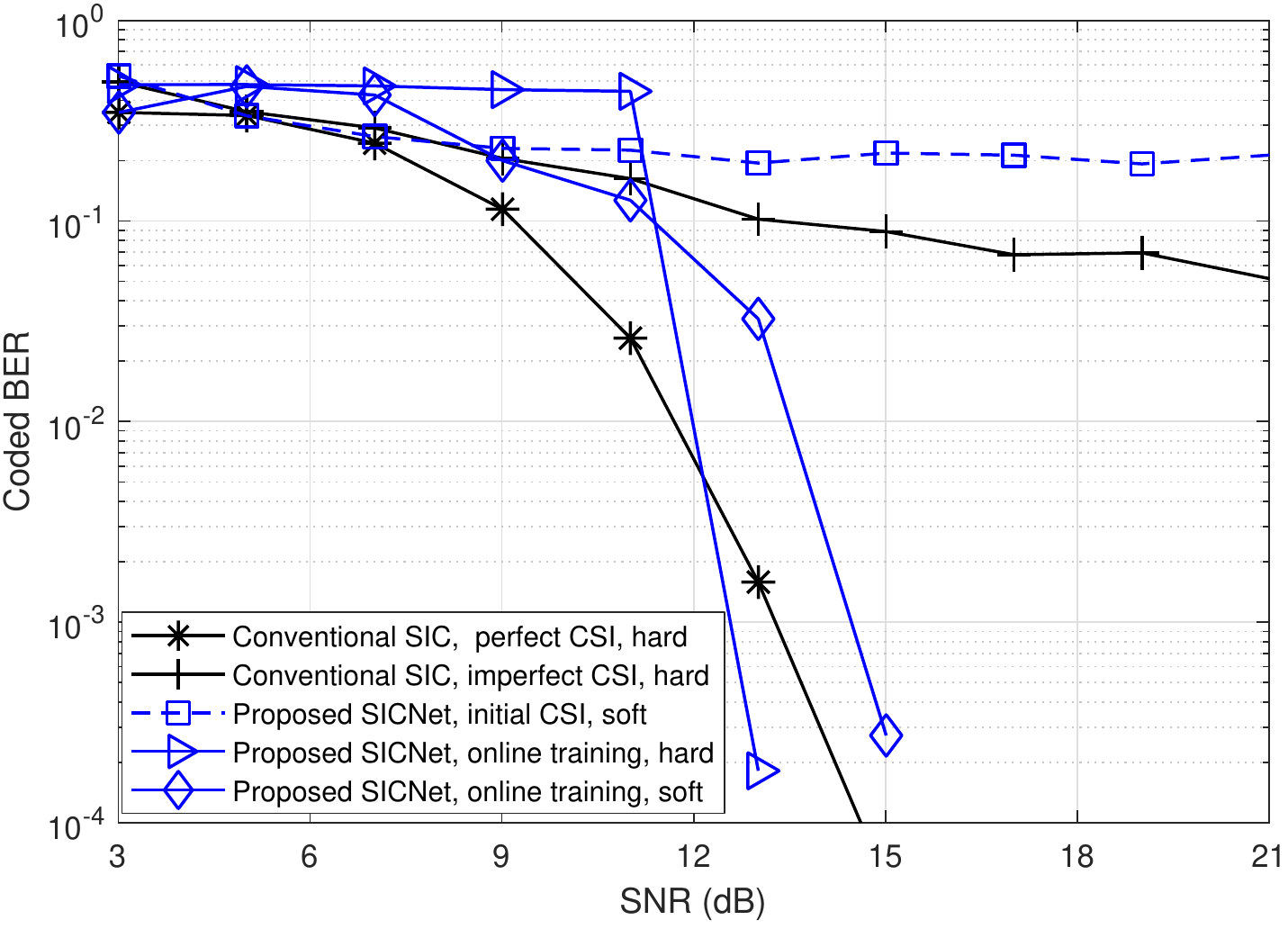} 
\par\end{centering}
\caption{Coded BER performance of SICNet with FEC-aided online training and
baselines under block fading channels and Gaussian noise. Here, the
proposed SICNet detector is trained with the \textit{local loss}.
\label{fig:fading}}
\end{figure}

Next, we demonstrate how the power of coded communications can be
exploited to facilitate online retraining of SICNet in block fading
channels. Fig \ref{fig:fading} illustrates the coded BER comparison
between our SICNet with FEC-aided online training and its baselines
under block fading channels and Gaussian noise. Here, the channel of
user 3 varies between blocks according to
$h_{3}\left(t\right)=0.8+\text{cos}\left(\frac{2\pi t}{17}\right)$,
where $t=0,1,...,99$ is the fading block index.
As such, there is a total of 100 fading blocks, each of which contains
1000 data bits, which produce 2000 uncoded bits when the 1/2
convolutional code $[7,5]$ is used. The classical SIC employs a hard
decoder to achieve better BER as shown in Fig. \ref{fig:cc_coded},
while SICNet uses two decoder types. Here, SICNet is initially trained
with 200 epochs over 5000 data samples of the first block with $t=0$,
and then the FEC-aided online training detailed in
Subsection~\ref{subsec:OnlineFEC} is performed using only 10 epochs
over 2000 online-recovered samples each following block. We also
include the BER of SICNet trained only with initial CSI, i.e., without
online training, for comparison. It is shown in Fig.~\ref{fig:fading}
that when the SNR is sufficiently high, i.e., $>12$ dB, SICNet with
online training approaches the performance of classical SIC, which
relies on perfect instantaneous CSI, whose performance is degraded
under imperfect CSI. Our SICNet using soft decoder even outperforms
its baseline with perfect CSI at SNRs in excess of 13
dB. \textcolor{black}{This benefit is substantial, since unlike the
classical scheme, our scheme does not suffer from any channel
estimation overhead, instead, it only involves a lightweight
re-training process relying on a few epochs.} Moreover, SICNet trained
with the initial CSI achieves poor coded BER performance, indicating
the importance of the proposed FEC-aided online training in order to
accurately track block fading channels. However,
observe from Fig.~\ref{fig:fading} that SICNet only performs well at
relatively high SNRs. At low SNRs one can utilize alternative
self-supervised learning strategies, such as the symbol-level
confidence approach proposed in \cite{Sun2021Generative}, or the
online training based on syndrome codes~\cite{teng2020syndrome}. We
set aside the joint study of SICNet with such strategies for our
future research. \textcolor{black}{Additionally, the benefits of SICNet can be exploited in a range of emerging scenarios such as short-packet communications \cite{Sun2018Short} and physical layer security \cite{Feng2019Beam}. We leave the study of such setups for our future work.} 

\vspace{-0.2cm}

\section{Conclusions\label{sec:Conclusions}}

\vspace{-0.1cm}
 In this paper, we proposed SICNet, which is a deep learning-aided
receiver for the downlink of non-orthogonal systems. In particular, SICNet uses DNNs
to replace the interference cancellation blocks of the model-based
SIC, where the soft information of each symbol is decoded by DNNs,
rather than by hard-decision ML detection. As a result, SICNet learns
to implement the model-based SIC in a data-driven manner, without
requiring any knowledge of channel models. Simulation results showed
that SICNet approaches the performance of the model-based SIC scheme
endowed with perfect CSI, and substantially outperforms its model-based counterpart
under CSI uncertainty, for both linear and non-linear channels. Additionally, SICNet is shown to be robust to  variations in the superposition code, and can reliably detect without reconstructing its architecture, and often even without retraining. It
was also demonstrated that SICNet learns to produce accurate LLRs,
leading to an improved performance over the model-based SIC,
when combined with soft FEC decoding. Finally, we  designed an FEC-aided
online training scheme for SICNet, which is capable of adapting to
the changes of block fading channels, achieving a BER performance close to
or even better than the model-based SIC employing perfect CSI at high SNRs,
particularly when soft decoder is used.

 \bibliographystyle{IEEEtran}
\bibliography{IEEEabrv,refs}

\begin{thebibliography}{10}
\providecommand{\url}[1]{#1}
\csname url@samestyle\endcsname
\providecommand{\newblock}{\relax}
\providecommand{\bibinfo}[2]{#2}
\providecommand{\BIBentrySTDinterwordspacing}{\spaceskip=0pt\relax}
\providecommand{\BIBentryALTinterwordstretchfactor}{4}
\providecommand{\BIBentryALTinterwordspacing}{\spaceskip=\fontdimen2\font plus
\BIBentryALTinterwordstretchfactor\fontdimen3\font minus
  \fontdimen4\font\relax}
\providecommand{\BIBforeignlanguage}[2]{{%
\expandafter\ifx\csname l@#1\endcsname\relax
\typeout{** WARNING: IEEEtran.bst: No hyphenation pattern has been}%
\typeout{** loaded for the language `#1'. Using the pattern for}%
\typeout{** the default language instead.}%
\else
\language=\csname l@#1\endcsname
\fi
#2}}
\providecommand{\BIBdecl}{\relax}
\BIBdecl

\bibitem{Liu2017SurveyNOMA}
Y.~{Liu}, Z.~{Qin}, M.~{Elkashlan}, Z.~{Ding}, A.~{Nallanathan}, and
  L.~{Hanzo}, ``Nonorthogonal multiple access for {5G} and beyond,''
  \emph{Proc. IEEE}, vol. 105, no.~12, pp. 2347--2381, 2017.

\bibitem{Ding2017SurveyNOMA}
Z.~{Ding}, X.~{Lei}, G.~K. {Karagiannidis}, R.~{Schober}, J.~{Yuan}, and V.~K.
  {Bhargava}, ``A survey on non-orthogonal multiple access for {5G} networks:
  Research challenges and future trends,'' \emph{IEEE J. Sel. Areas Commun.},
  vol.~35, no.~10, pp. 2181--2195, 2017.

\bibitem{andrews2005interference}
J.~G. Andrews, ``Interference cancellation for cellular systems: a contemporary
  overview,'' \emph{{IEEE} Wireless Commun.}, vol.~12, no.~2, pp. 19--29, 2005.

\bibitem{Yang2016imcsi}
Z.~{Yang}, Z.~{Ding}, P.~{Fan}, and G.~K. {Karagiannidis}, ``On the performance
  of non-orthogonal multiple access systems with partial channel information,''
  \emph{IEEE Trans. Commun.}, vol.~64, no.~2, pp. 654--667, 2016.

\bibitem{shlezinger2018asymptotic}
N.~Shlezinger, Y.~C. Eldar, and M.~R. Rodrigues, ``Asymptotic task-based
  quantization with application to massive {MIMO},'' \emph{{IEEE} Trans. Signal
  Process.}, vol.~67, no.~15, pp. 3995--4012, 2019.

\bibitem{iofedov2015mimo}
I.~Iofedov and D.~Wulich, ``{MIMO--OFDM} with nonlinear power amplifiers,''
  \emph{{IEEE} Trans. Commun.}, vol.~63, no.~12, pp. 4894--4904, 2015.

\bibitem{kobayashi2001successive}
M.~Kobayashi, J.~Boutros, and G.~Caire, ``Successive interference cancellation
  with {SISO} decoding and {EM} channel estimation,'' \emph{{IEEE} J. Sel.
  Areas Commun.}, vol.~19, no.~8, pp. 1450--1460, 2001.

\bibitem{khani2020adaptive}
M.~Khani, M.~Alizadeh, J.~Hoydis, and P.~Fleming, ``Adaptive neural signal
  detection for massive {MIMO},'' \emph{{IEEE} Trans. Wireless Commun.},
  vol.~19, no.~8, pp. 5635--5648, 2020.

\bibitem{Wang2020survey}
J.~{Wang}, C.~{Jiang}, H.~{Zhang}, Y.~{Ren}, K.~C. {Chen}, and L.~{Hanzo},
  ``Thirty years of machine learning: The road to {Pareto}-optimal wireless
  networks,'' \emph{{IEEE} Commun. Surveys Tuts.}, vol.~22, no.~3, pp.
  1472--1514, 2020.

\bibitem{farsad2020data}
N.~Farsad, N.~Shlezinger, A.~J. Goldsmith, and Y.~C. Eldar, ``Data-driven
  symbol detection via model-based machine learning,'' \emph{Communications in
  Information and Systems}, vol.~20, no.~3, pp. 283--317, 2020.

\bibitem{Bengio09learning}
Y.~Bengio, ``Learning deep architectures for {AI},'' \emph{Foundations and
  Trends in Machine Learning}, vol.~2, no.~1, pp. 1--127, 2009.

\bibitem{Luong2020MC-AE}
T.~V. {Luong}, Y.~{Ko}, M.~{Matthaiou}, N.~A. {Vien}, M.~T. {Le}, and V.~D.
  {Ngo}, ``Deep learning-aided multicarrier systems,'' \emph{IEEE Trans.
  Wireless Commun.}, vol.~20, no.~3, pp. 2109--2119, 2021.

\bibitem{Alberge2018Constell}
F.~{Alberge}, ``Constellation design with deep learning for downlink
  non-orthogonal multiple access,'' in \emph{Proc IEEE PIMRC}, 2018, pp. 1--5.

\bibitem{Kang2020LearningSIC}
J.~M. {Kang}, I.~M. {Kim}, and C.~J. {Chun}, ``Deep learning-based {MIMO-NOMA}
  with imperfect {SIC} decoding,'' \emph{IEEE Syst. J.}, vol.~14, no.~3, pp.
  3414--3417, 2020.

\bibitem{Thien2020energy}
T.~V. {Luong}, Y.~{Ko}, N.~A. {Vien}, M.~{Matthaiou}, and H.~Q. {Ngo}, ``Deep
  energy autoencoder for noncoherent multicarrier {MU-SIMO} systems,''
  \emph{IEEE Trans. Wireless Commun.}, vol.~19, no.~6, pp. 3952--3962, 2020.

\bibitem{Luong2018impact}
T.~V. {Luong} and Y.~{Ko}, ``Impact of {CSI} uncertainty on {MCIK-OFDM}: Tight
  closed-form symbol error probability analysis,'' \emph{IEEE Trans. Veh.
  Technol.}, vol.~67, no.~2, pp. 1272--1279, 2018.

\bibitem{Luong2018spread}
T.~V. Luong and Y.~Ko, ``Spread {OFDM-IM} with precoding matrix and
  low-complexity detection designs,'' \emph{IEEE Trans. Veh. Technol.},
  vol.~67, no.~12, pp. 11\,619--11\,626, 2018.

\bibitem{Shlezinger2020ViterbiNet}
N.~{Shlezinger}, N.~{Farsad}, Y.~C. {Eldar}, and A.~J. {Goldsmith},
  ``Viterbi{N}et: A deep learning based {Viterbi} algorithm for symbol
  detection,'' \emph{IEEE Trans. Wireless Commun.}, vol.~19, no.~5, pp.
  3319--3331, May 2020.

\bibitem{shlezinger2020data}
N.~Shlezinger, N.~Farsad, Y.~C. Eldar, and A.~J. Goldsmith, ``Data-driven
  factor graphs for deep symbol detection,'' in \emph{Proc. IEEE ISIT}, 2020,
  pp. 2682--2687.

\bibitem{Shlezinger2020DeepSIC}
N.~{Shlezinger}, R.~{Fu}, and Y.~C. {Eldar}, ``{DeepSIC}: Deep soft
  interference cancellation for multiuser {MIMO} detection,'' \emph{{IEEE}
  Trans. Wireless Commun.}, vol.~20, no.~2, pp. 1349--1362, 2021.

\bibitem{shlezinger2020model}
N.~Shlezinger, J.~Whang, Y.~C. Eldar, and A.~G. Dimakis, ``Model-based deep
  learning,'' \emph{arXiv preprint arXiv:2012.08405}, 2020.

\bibitem{choi2000iterative}
W.-J. Choi, K.-W. Cheong, and J.~M. Cioffi, ``Iterative soft interference
  cancellation for multiple antenna systems.'' in \emph{Proc. IEEE WCNC}, 2000,
  pp. 304--309.

\bibitem{teng2020syndrome}
C.-F. Teng and Y.-L. Chen, ``Syndrome-enabled unsupervised learning for neural
  network-based polar decoder and jointly optimized blind equalizer,''
  \emph{{IEEE} Trans. Emerg. Sel. Topics Circuits Syst.}, vol.~10, no.~2, pp.
  177--188, 2020.

\bibitem{raviv2021meta}
T.~Raviv, S.~Park, N.~Shlezinger, O.~Simeone, Y.~C. Eldar, and J.~Kang,
  ``Meta-{V}iterbi{N}et: Online meta-learned {V}iterbi equalization for
  non-stationary channels,'' in \emph{Proc. IEEE ICC Workshops}, 2021, pp.
  1--6.

\bibitem{dahlman20103g}
E.~Dahlman, S.~Parkvall, J.~Skold, and P.~Beming, \emph{{3G} evolution: {HSPA}
  and {LTE} for mobile broadband}.\hskip 1em plus 0.5em minus 0.4em\relax
  Academic press, 2010.

\bibitem{Yuan2018codedSIC}
L.~{Yuan}, J.~{Pan}, N.~{Yang}, Z.~{Ding}, and J.~{Yuan}, ``Successive
  interference cancellation for {LDPC} coded nonorthogonal multiple access
  systems,'' \emph{IEEE Trans. Veh. Technol.}, vol.~67, no.~6, pp. 5460--5464,
  2018.

\bibitem{alexandropoulos2021reconfigurable}
G.~C. Alexandropoulos, N.~Shlezinger, and P.~del Hougne, ``Reconfigurable
  intelligent surfaces for rich scattering wireless communications: Recent
  experiments, challenges, and opportunities,'' \emph{{IEEE} Commun. Mag.},
  vol.~59, no.~6, pp. 28--35, 2021.

\bibitem{goodfellow2016deep}
I.~Goodfellow, Y.~Bengio, and A.~Courville, \emph{Deep learning}.\hskip 1em
  plus 0.5em minus 0.4em\relax MIT press, 2016.

\bibitem{park2020meta}
S.~Park, H.~Jang, O.~Simeone, and J.~Kang, ``Learning to demodulate from few
  pilots via offline and online meta-learning,'' \emph{IEEE Trans. Signal
  Proces.}, vol.~69, pp. 226--239, 2020.

\bibitem{raviv2020data}
T.~Raviv, N.~Raviv, and Y.~Be'ery, ``Data-driven ensembles for deep and
  hard-decision hybrid decoding,'' in \emph{Proc. IEEE ISIT}, 2020, pp.
  321--326.

\bibitem{Sun2021Generative}
L.~{Sun}, Y.~{Wang}, A.~L. {Swindlehurst}, and X.~{Tang},
  ``Generative-adversarial-network enabled signal detection for communication
  systems with unknown channel models,'' \emph{{IEEE} J. Sel. Areas Commun.},
  vol.~39, no.~1, pp. 47--60, 2021.

\bibitem{finish2022symbol}
R.~A. Finish, Y.~Cohen, T.~Raviv, and N.~Shlezinger, ``Symbol-level online
  channel tracking for deep receivers,'' in \emph{Proc. IEEE ICASSP}, 2022.

\bibitem{DeepIM2019}
T.~V. {Luong}, Y.~{Ko}, N.~A. {Vien}, D.~H.~N. {Nguyen}, and M.~{Matthaiou},
  ``Deep learning-based detector for {OFDM-IM},'' \emph{IEEE Wireless Commun.
  Lett.}, vol.~8, no.~4, pp. 1159--1162, Aug. 2019.

\bibitem{Kingma2014AdamAM}
D.~P. Kingma and J.~Ba, ``Adam: A method for stochastic optimization,''
  \emph{ICLR}, pp. 1--15, 2014.

\bibitem{jeon2018supervised}
Y.-S. Jeon, S.-N. Hong, and N.~Lee, ``Supervised-learning-aided communication
  framework for {MIMO} systems with low-resolution {ADCs},'' \emph{IEEE Trans.
  Veh. Technol.}, vol.~67, no.~8, pp. 7299--7313, 2018.

\bibitem{Sun2018Short}
X.~Sun, S.~Yan, N.~Yang, Z.~Ding, C.~Shen, and Z.~Zhong, ``Short-packet
  downlink transmission with non-orthogonal multiple access,'' \emph{{IEEE}
  Trans. Wireless Commun.}, vol.~17, no.~7, pp. 4550--4564, 2018.

\bibitem{Feng2019Beam}
Y.~Feng, S.~Yan, Z.~Yang, N.~Yang, and J.~Yuan, ``Beamforming design and power
  allocation for secure transmission with noma,'' \emph{{IEEE} Trans. Wireless
  Commun.}, vol.~18, no.~5, pp. 2639--2651, 2019.

\end{thebibliography}

\end{document}